\documentclass[aps, prd,twocolumn,superscriptaddress,nofootinbib,preprintnumbers]{revtex4-1}
\usepackage[utf8]{inputenc}
\usepackage[english]{babel}
\usepackage{natbib}
\usepackage{graphicx}
\usepackage{amsmath,amssymb}
\usepackage{hyperref}
\usepackage{braket}
\usepackage{float}
\usepackage[dvipsnames]{xcolor}
\usepackage{soul}
\usepackage{rotating}
\usepackage{multirow}
\usepackage{mathtools}
\usepackage{makecell}

\usepackage[caption = false]{subfig}
\usepackage{txfonts}
\usepackage[margin=0.75in]{geometry}
\usepackage{tabularx}

\hypersetup{
	colorlinks=true,
	linkcolor=red,
	citecolor=blue,
}

\usepackage{xcolor}

\DeclareMathAlphabet\mathbfcal{OMS}{cmsy}{b}{n}

\newcommand{\beq}{\begin{equation}}
\newcommand{\eeq}{\end{equation}}
\newcommand{\beqa}{\begin{eqnarray}}
\newcommand{\eeqa}{\end{eqnarray}}

\definecolor{darkgreen}{rgb}{0.0, 0.5, 0.0}
\definecolor{darkcyanxf}{RGB}{0.0, 139.0, 139.0}

\newcommand{\msun}[1]{$M_{\odot}$}

\newcommand{\avg}[1]{\langle #1 \rangle}

\newcommand{\mcal}{\textsc{metacalibration}}

\newcommand{\xigty}{\mbox{$\xi_{\gamma_{t} y}$}}
\defcitealias{Battaglia:2012}{B12}

\defcitealias{paperB}{paper II}
\newcommand{\paperB}{\citetalias{paperB}}

\usepackage{lineno}
\begin{document}
\title[shear x y ]{Cross-correlation of DES Y3 lensing  and \textsc{ACT}/\textit{Planck} thermal Sunyaev Zel'dovich Effect I: Measurements, systematics tests, and feedback model constraints
}

% \date{Last updated \today}

\label{firstpage}

\begin{abstract}
We present a tomographic measurement of the cross-correlation between thermal Sunyaev-Zeldovich (tSZ) maps from ${\it Planck}$ and the Atacama Cosmology Telescope (ACT) and weak galaxy lensing shears measured during the first three years of observations of the Dark Energy Survey (DES Y3). {This correlation is sensitive to the thermal energy in baryons over a wide redshift range, and is therefore a powerful probe of astrophysical feedback.}  We detect the {correlation} at a statistical significance of $21\sigma$, the highest significance {to date}. {We examine the tSZ maps for potential contaminants, including  cosmic infrared background (CIB) and radio sources, finding that CIB has a substantial impact on our measurements and must be taken into account in our analysis}. We use the cross-correlation measurements to test different feedback models. 
In particular, we model the tSZ using several different pressure profile models calibrated against hydrodynamical simulations.
Our analysis marginalises over redshift uncertainties, shear calibration biases, and intrinsic alignment effects. We also marginalise over $\Omega_{\rm m}$ and $\sigma_8$ using ${\it Planck}$ or DES priors. We find that the data prefers the model with a low amplitude of the pressure profile at small scales, compatible with a scenario with strong AGN feedback and ejection of gas from the inner part of the halos. When using a more flexible model for the shear profile, constraints are weaker, and the data cannot discriminate between different baryonic prescriptions.
\end{abstract}
\preprint{DES-21-333}
\preprint{FERMILAB-PUB-21-333-AE}

\author{M.~Gatti}\email{marcogatti29@gmail.com}
\affiliation{Department of Physics and Astronomy, University of Pennsylvania, Philadelphia, PA 19104, USA}
\author{S.~Pandey}
\affiliation{Department of Physics and Astronomy, University of Pennsylvania, Philadelphia, PA 19104, USA}
\author{E.~Baxter}
\affiliation{Institute for Astronomy, University of Hawai'i, 2680 Woodlawn Drive, Honolulu, HI 96822, USA}
\author{J.~C.~Hill}
\affiliation{Department of Physics, Columbia University, New York, NY, USA 10027}
\affiliation{Center for Computational Astrophysics, Flatiron Institute, New York, NY, USA 10010}
\author{E.~Moser}
\affiliation{Department of Astronomy, Cornell University, Ithaca, NY 14853, USA}
\author{M.~Raveri}
\affiliation{Department of Physics and Astronomy, University of Pennsylvania, Philadelphia, PA 19104, USA}
\author{X.~Fang}
\affiliation{Department of Astronomy/Steward Observatory, University of Arizona, 933 North Cherry Avenue, Tucson, AZ 85721-0065, USA}
\author{J.~DeRose}
\affiliation{Lawrence Berkeley National Laboratory, 1 Cyclotron Road, Berkeley, CA 94720, USA}
\author{G.~Giannini}
\affiliation{Institut de F\'{\i}sica d'Altes Energies (IFAE), The Barcelona Institute of Science and Technology, Campus UAB, 08193 Bellaterra (Barcelona) Spain}
\author{C.~Doux}
\affiliation{Department of Physics and Astronomy, University of Pennsylvania, Philadelphia, PA 19104, USA}
% ...
\collaboration{DES and ACT Collaboration}

%\begin{keywords}
%gravitational lensing: weak - methods: data analysis - %techniques: image processing - catalogues - surveys - cosmology: %observations.
%\end{keywords}

\maketitle


%\blfootnote{$^{\star}$ E-mail: mgatti@ifae.es}
%\blfootnote{$^{\dag}$ E-mail: erin.sheldon@gmail.com}

%%%%%%%%%%%%%%%%%%%%%%%%%%%%%%%%%%%%%%%%%%%%%%%%%%PSFre
%%%%%%%%%%%%%%%%% BODY OF PAPER %%%%%%%%%%%%%%%%%%

\section{Introduction}

%Not only does the Cosmic Microwave Background (CMB) represent an excellent laboratory to study the early Universe physics, but it can also be used to effectively probe the properties of the late Universe. 
The Cosmic Microwave Background (CMB) provides a means to study early Universe physics as well as a powerful tool with which to probe the properties of the late Universe.
As photons travel through cosmic time, they are affected by the large-scale structure of the Universe at low redshift, which leaves an imprint on the CMB. Among these so called ``secondary anisotropies,’’ generated after photons leave the surface of last scattering, the imprints left by the thermal Sunyaev–Zel’dovich (tSZ) effect \citep{Sunyaev1972,Sunyaev1980} are some of the most important. The effect is caused by inverse Compton scattering of CMB photons with ionised gas. {The tSZ effect is an effective probe of large-scale structure, as the signal is sensitive to the halo mass function, which in turn strongly depends on the amplitude of the matter fluctuations, i.e., $\sigma_8$, and on the matter density, $\Omega_{\rm m}$ \citep{komatsu2002}}. It is also an effective probe of the properties of the hot gas within and outside dark matter halos, as the measured signal depends on the hot gas pressure profile. 

A better understanding of the properties of baryons within dark matter halos is needed to fully exploit the cosmological information from the small-scale regime in current and future cosmological analyses (DES, \citet{Flaugher2015}; KIDS, \citet{Kuijken2015}; HSC, \citet{Aihara2018}; Rubin LSST, \citet{Abell2009}; Euclid, \citet{Laureijs2011}). Astrophysical feedback significantly impacts the baryons, leading to changes in the matter power spectrum at small scales \citep{vanDaalen2011,Chisari2018,vanDaalen2020}.  Ignoring such effects can lead to significant biases in cosmological analyses \citep{Semboloni2013,Eifler2015,Huang2019}.

Various strategies have been adopted to mitigate the impact of baryonic feedback on cosmological constraints. The most straightforward way is to exclude the scales that could be significantly affected by baryonic effects (e.g., \citet{Troxel2018}). %by comparing with the expected contribution from different hydrodynamical simulations. 
Other methods include adding extra complexity to the modelling to account for the effect of baryons \citep{Mead2016,Mead2020,Joudaki2017, Hildebrandt2017,Asgari2020}, empirically modelling baryonic effects using fitting formulae calibrated against hydrodynamical simulations \citep{Hikage2019}, or using principal component analysis and hydrodynamical simulations to identify the modes of the data vector most sensitive to baryonic effects, and to marginalise over them \citep{Eifler2015,Huang2019,Huang2020}.  As many of these mitigation strategies rely more or less directly on hydrodynamical simulations, the specific details of the implementation of baryonic physics in such simulations also have an impact on these methods.

Analysis of the tSZ effect provides a potential means for setting priors on different baryonic feedback prescriptions, or to promote or rule out some of the hydrodynamical simulations. %.used to calibrate the aforementioned methods.
Particularly appealing are studies that involve the cross-correlation of the tSZ effect with other probes sensitive to large-scale structure.  Such cross-correlations have different sensitivity to nuisance parameters that makes these measurements less prone to systematics. Moreover, cross-correlations with different probes are key to study the evolution of baryonic effects with redshift or their dependence on the environment and the halo mass. In this work we focus on cross-correlations between the tSZ effect and weak gravitational lensing, a measurement that has gained attention over the last few years. \citet{vW2014} obtained the first detection of the cross correlation signal between the shear field and a Compton-$y$ measurement, using a Canada France Hawaii Telescope (CFHT) lensing convergence map and a {Compton-$y$ map built using \textit{Planck} data}. In the following year, \citet{Hill2014} measured the tSZ x CMB lensing signal for the first time, which is in spirit a similar measurement, although it probes a higher redshift range compared to \cite{vW2014}. Subsequently, other measurements have been performed by \citet{Hojjati20117}, who detected a cross-correlation signal using \textit{Planck} data and a shape catalog from the RCSLenS survey, and by \citet{Osato2020} using \textit{Planck} and HSC data. For this work, we use the fiducial shape catalog for the DES Y3 data \citep{y3-shapecatalog}, and Compton-$y$ maps from both \textit{Planck} \citep{Planck_2016_tsz} and the Atacama Cosmology Telescope ACT \citep{Madhavacheril2020}.\footnote{The ACT Compton-$y$ map is created using both low (spatial) resolution data from \textit{Planck} and high resolution data from ACT, but for the sake of simplicity we refer to it as the ACT map. More details are given in \S~\ref{sect:data}. } The large area coverage by the DES Y3 weak lensing sample (4183 sq. degrees) allows us to considerably improve the signal-to-noise of the measurement compared to previous studies. Moreover, the addition of the ACT map --- which covers a smaller area compared to \textit{Planck} but has a much higher spatial resolution --- allows us to extend the measurement down to $\sim$ 2.5 arcminutes scales.

In this work and in a companion paper (\citet{paperB}, hereafter paper II) we present the correlation measurements and perform several different analyses.  We focus here on two aspects in particular:
\begin{itemize}
    \item we discuss various systematic tests, with a focus on the effect of potential contaminants, i.e. the cosmic infrared background (CIB), and radio sources;
    \item we compare the measurements to theoretical predictions using the halo model framework and pressure profiles as estimated from a number of hydrodynamical simulations, with the goal of discriminating between different baryonic feedback models.
\end{itemize}
The analysis performed in this paper treats the pressure profile predictions of hydrodynamical simulations as fixed, and fits the data marginalizing over several nuisance parameters, modelling astrophysical and measurement systematics, including photometric redshift uncertainties, intrinsic alignment and shear calibration biases. We also marginalise over cosmological parameters assuming \textit{Planck} or DES priors. On the other hand, in \paperB, we use an alternate approach:
\begin{itemize}
    \item we fit the measurements by varying the parameters of a flexible model for the halo pressure profiles, exploring how the halo pressure profiles evolve as a function of halo mass and redshift;
    \item we discuss implications of our measurements on the constraints of the so-called halo mass bias parameter.
\end{itemize}

%In this work we perform a number of systematic tests. In particular, we test that our measurement is robust against a number of potential contaminants: the cosmic infrared background (CIB), and radio sources. After characterising the measurement, we compare it to theoretical predictions using the halo model framework and pressure profiles as estimated from a number of hydrodynamical simulations. In the comparison, we marginalise over a number of nuisance parameters modelling astrophysical and measurement systematics, including photometric redshift uncertainties, intrinsic alignment and shear calibration biases. We assume both \textit{Planck} and DES priors for the cosmological parameters, and we quantify the level of agreement/disagreement of each model with the data using Gaussian estimators \citep{Raveri2019}.

%The approach that we adopt in this paper is to treat the pressure profile predictions of hydrodynamical simulations as fixed, and to then fit the data marginalizing over several nuisance parameters.  This allows us to determine where the simulation predictions are consistent with the data using goodness of fit tests.  An alternate approach, that we adopt in a companion paper (\citet{paperB}, hereafter paper II), is to instead fit the data varying the parameters of the pressure profile models.  These constraints can the be translated into quantities such as the relationship between integrated tSZ signal and halo mass.

The paper is organised as follows. In \S~\ref{sect:data}
 we describe the data used in this work. \S~\ref{sect:modeling}
 describes the theoretical modelling of the measurement, introducing the feedback models considered in this work and the modelling choices of the analysis. \S~\ref{sect:measurement} presents our \textit{Planck} x DES and ACT x DES measurements, and systematic tests are discussed in \S~\ref{sect:map_tests}. We test different feedback models in \S~\ref{sect:feedback_models}; we summarise our findings in \S~\ref{sec:summary}. We provide further validation of our modelling on N-body simulations in Appendix \ref{sect:sims_buzzard}; Appendix \ref{sec:covariance} shows our validation of the analytical covariance matrix; Appendix \ref{sect:weird_sect} illustrates the effect of CIB contamination on simulated Compton-$y$ maps;  last, Appendix \ref{sec:DES_prior} shows our results when DES priors for the cosmological parameters are assumed.

\section{Data Products}\label{sect:data}

\begin{figure*}
\includegraphics[width=1.\textwidth]{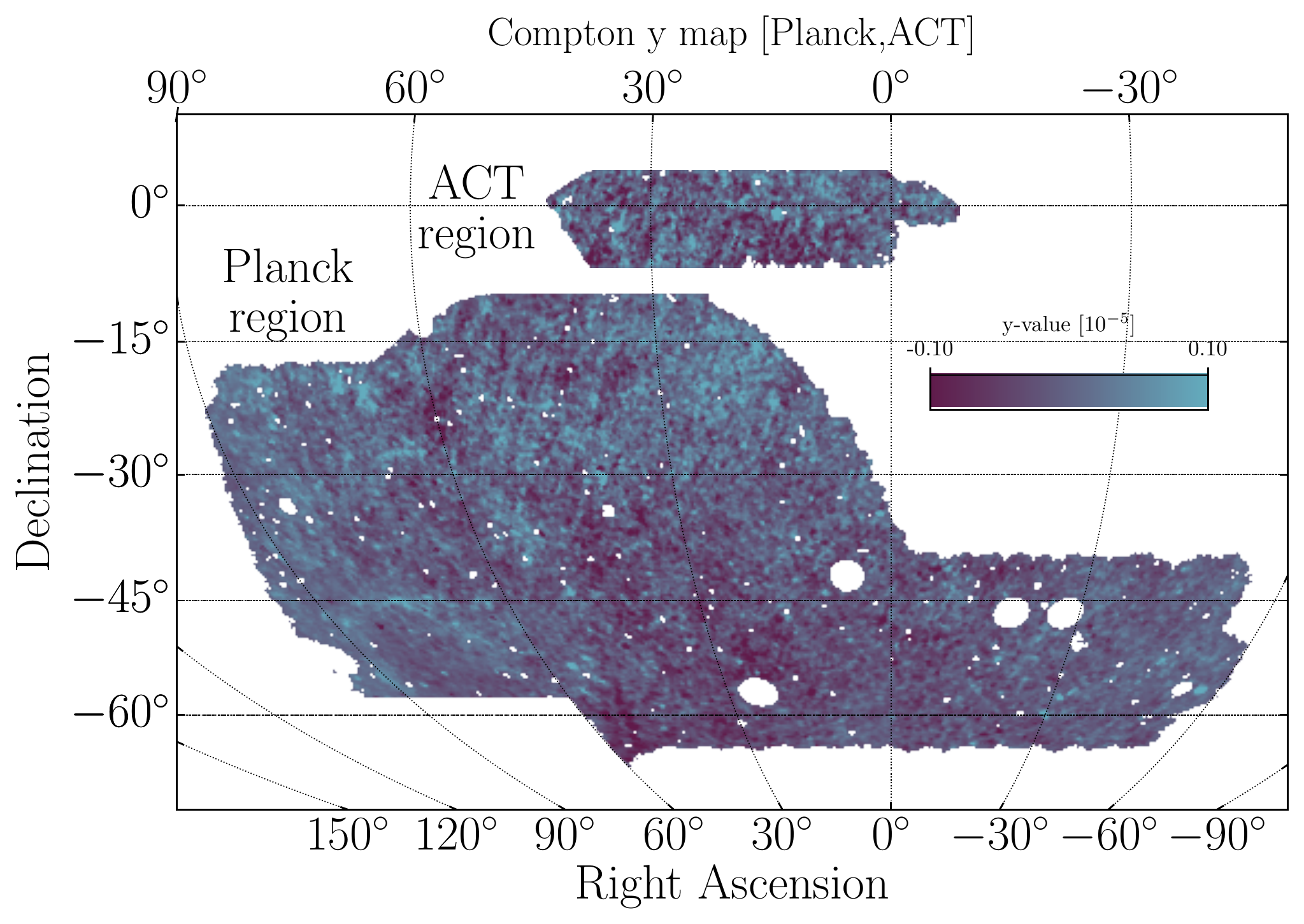}
\caption[]{\textit{Planck} and ACT Compton y-maps after isolating the part of the maps that overlap with the DES Y3 footprint. A small ``buffer'' region between the two maps has been removed to reduce correlation between measurements in the two patches. }
\label{fig:y_fp}
\end{figure*}

%In this work, we focus on the cross correlation between the fiducial DES Y3 shape catalog and two different Compton-y maps, estimated from \textit{Planck} and ACT respectively. \textit{Planck} and ACT maps differ in both spatial coverage and resolution, as explained below.

\subsection{Planck Compton map}

We use the publicly available 2015 \textit{Planck} High Frequency Instrument (HFI) and Low Frequency Instrument (LFI) maps \citep{Planck_LOW,Planck_HIGH} to estimate the Compton-$y$ map using the Needlet Internal Linear Combination (NILC) algorithm \citep{Guilloux2007,Delabrouille2009}. We build our own version of the \textit{Planck} Compton-$y$ map, also using different prescriptions to de-project (i.e., remove) the contamination by the cosmic infrared background (CIB).

In particular, we use all the channels from 30 GHz to 545 GHz. We do not include the frequency map at 857 GHz because (1) the calibration of this map is more uncertain than the other frequency maps; (2) the dust is much brighter in this map than in any of the other maps and there are large dust-related residuals found in NILC maps if the 857 GHz map is used \citep{Planck_2016_tsz}. We also estimate the Compton-$y$ map excluding the 545 GHz channel {to test the sensitivity of our results to the CIB, which is brightest at high frequencies}. The details of the implementation of this algorithm are presented in Appendix A of \citet{Pandey:2019} and more details on the CIB contribution are given in \S~\ref{sect:map_tests}. {We found that the signal obtained using our own version of the Compton-$y$ map with no CIB de-projection was compatible with the signal obtained using the public \textit{Planck} Compton-$y$ map \citep{Planck_2016_tsz}.} The maps come in \texttt{HEALPIX} format with a resolution of \texttt{NSIDE} = 2048. The \textit{Planck} $y$-map resolution has an effective Full-Width-Half-Maximum (FWHM) of 10 arcminutes. {When producing the Compton-$y$ maps, we applied the standard \textit{Planck} foreground mask, which limits the diffuse Galactic emission removing the most-contaminated $\sim$40 per cent of the sky, mostly around the Galactic plane. We further applied the fiducial DES Y3 mask \citep{y3-gold}, which removes areas affected by astrophysical foregrounds (e.g., bright stars and large nearby galaxies) and ‘bad’ regions with recognised data processing issues within the DES Y3 footprint. In the fiducial maps we did not mask radio sources; however, for the purpose of testing, we produced an alternate version of the maps by removing the pixels affected by radio sources detected by ACT (which detects sources to a fainter limit compared to \textit{Planck}).} Lastly, we removed the regions that have overlap with ACT data, since that part of the sky is covered by the ACT + \textit{Planck} Compton-$y$ map (described in the next section) and the latter is preferred as it comes at a higher spatial resolution. In order to avoid correlations between the two Compton-$y$ maps during the analysis, we further cut out a ``buffer'' region of a few degrees of width between the two maps, as shown in Fig. \ref{fig:y_fp}. This reduces the covariance between the measurements obtained using the two maps. The final area covered by \textit{Planck} data is 3423 square degrees.

%In the case of NILC we also use the LFI data at large angular
%%scales (` < 300). Similarly, for both methods the 857 GHz map,
%which traces the thermal dust emission on large angular scales, is

 %We use the \texttt{NILC} ( Compton-$y$ map from \textit{Planck} \citep{Planck_2016_tsz}, obtained using \sout{combining} frequency maps from 30 to 857 GHz 
 %and a component separation method based on the internal linear combination (ILC) approach. 
 
%For the purpose of testing, we also make use of a version of the \textit{Planck} map where relevant foreground ratio sources have been masked out (see \S~\ref{sect:map_tests}). In the same section, we test different versions of the \textit{Planck} Compton-$y$ map, obtained with a slightly different algorithm (called \texttt{NILC}, and with different potential contaminants de-projected). 

\subsection{ACT Compton map}

%One distinction is that our ILC weights include anisotropic dependence in the Fourier domain, which is not considered in NILC.

We use Compton-$y$ maps from the Atacama Cosmology Telescope (ACT), as presented in \cite{Madhavacheril2020} as part of DR4. The maps are obtained by combining \textit{Planck} maps from 30 to 545 GHz and ACT maps at 98 and 150 GHz, using an anisotropic ILC component separation approach in the 2D Fourier domain (slightly different from the one used to create \textit{Planck} maps). The original ACT maps are converted to \texttt{HEALPIX} format using the \textsc{pixell} package\footnote{\url{https://github.com/simonsobs/pixell}}; the maps have a resolution of \texttt{NSIDE} = 8192, with a FWHM of 1.6 arcminutes. We only use data from the D56 region, which overlaps with DES data, for a total area of 394 square degrees, after applying the ACT and DES masks. {In contrast to the \textit{Planck} maps case, for ACT compact sources (like radio sources) are subtracted by default, and the subtraction is followed by an inpainting procedure that estimates the correct value of the pixels affected by the compact sources. The inpainting algorithm fills holes around compact sources by finding the maximum-likelihood solution for pixels within the hole constrained by the pixels in a context region around the hole \citep{Madhavacheril2020}.} For the purpose of testing, we also make use of  versions of the ACT map with the CIB contribution de-projected (see \S~\ref{sect:map_tests}). Contrary to the \textit{Planck} map we do not create a version of the ACT Compton-$y$ map excluding the 545 GHz frequency channel, as the fiducial ACT maps already assign a very small weight to this frequency channel.

\subsection{DES Y3 Data}\label{sect:desdata}
\begin{figure}
\includegraphics[width=0.45\textwidth]{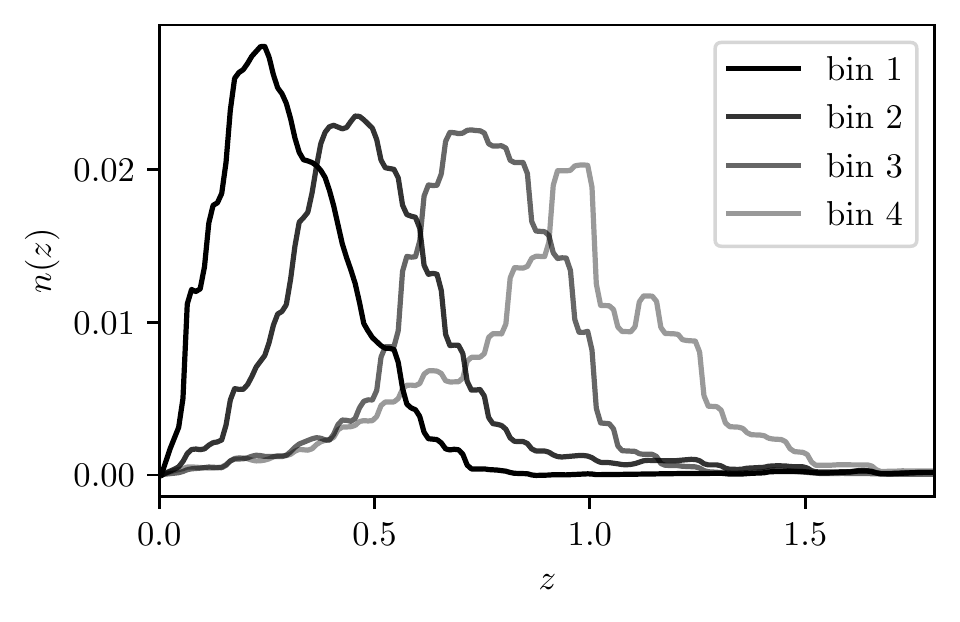}
\caption{Redshift distributions for the four DES Y3 tomographic bins \citep{y3-sompz}.}
\label{fig:nz_comp}
\end{figure}

We use the fiducial DES Y3 shape catalog, presented in \cite{y3-shapecatalog}.  The DES Y3 shape catalogue is created using the \mcal\ pipeline, which is able to self-calibrate the measured shapes against shear and selection biases by measuring the mean shear and selection response matrix of the sample. The current DES Y3 implementation of \mcal\ \citep{HuffMcal2017,SheldonMcal2017} is able to correct for shear biases up to a multiplicative factor of 2-3 per cent, which is fully characterised using image simulations \citep{y3-imagesims}. The final sample comprises 100 million objects, for an effective number density of $n_{\rm eff} = 5.59$ gal/arcmin$^{2}$, spanning an effective area of 4139 square degrees. Galaxies are further divided into 4 tomographic bins and redshift estimates for each of the tomographic bins are provided by the SOMPZ method \citep{y3-sompz}. {The method uses additional information from deep fields \cite{y3-deepfields} and spectroscopic samples to break degeneracies in the photo-$z$ estimates of the wide field; this is achieved by creating self-organising maps (SOMs) of the spectroscopic, the deep and wide field galaxies and mapping the three together. The redshift bin edges of the tomographic bins used for the tomographic bin assignments are z = [0.0, 0.358, 0.631, 0.872, 2.0]; wide field galaxies are assigned to different tomographic bins depending on the mean redshift of the cell of the deep SOM they are associated to. This assignment procedure, however, does not guarantee that the redshift distributions are sharply bounded (Fig. \ref{fig:nz_comp})} Throughout this paper, we use the fiducial DES Y3 priors for the shear calibration biases and redshift uncertainties (see \S~\ref{sect:like}).

\section{Theory}\label{sect:modeling}

There exist different tSZ-weak lensing cross-correlation estimators in the literature, both in configuration and harmonic space (for a discussion on different estimators see, e.g., \cite{Hojjati20117}). Some of these estimators require construction of a weak lensing convergence map from shear field estimates as an intermediate step, which usually requires extra care in the modelling of the signal due to systematics and uncertainties introduced by the map reconstruction process. In this work, however, we focus on a configuration space estimator, the $\xi^{\gamma_t y}$ correlation function, which does not require this intermediate step. Given a catalog of galaxy shapes and a Compton-$y$ map, such an estimator is constructed by measuring the tangential shear around every point of the $y$ map, for different angular separations $\theta$, and then averaging all the measurements. This estimator has the advantage of being particularly robust to additive systematics in the shear data.  The measured $\xi^{\gamma_t y}$ correlation signal can be theoretically modelled relying on the halo model framework \citep{Cooray2002}. {Note that the validity of the halo model to describe shear-Compton-$y$ cross-correlation measurements has been demonstrated by \cite{Battaglia2015}, using hydrodynamical simulations \citep{Battaglia_2012}.}

We begin by modelling correlations between the convergence field and Compton-$y$ maps in harmonic space, and then transform this model to obtain a prediction for the $\xi^{\gamma_t y}$ correlation signal in configuration space. In harmonic space, the correlation can be described as an effective sum of a one-halo term and a two-halo term, with the one-halo term given by an integral over redshift ($z$) and halo mass ($M$):
\begin{multline}\label{eq:Cl1h}
C_{\ell}^{\kappa,y; 1h} = \int_{z_{\rm{min}}}^{z_{\rm{max}}} dz \frac{dV}{dz d\Omega} \int_{M_{\rm{min}}}^{M_{\rm{max}}} dM \frac{dn}{dM} \bar{\kappa}_{\ell}(M,z) \ \bar{y}_{\ell}(M,z),
\end{multline}
where $dV$ is the cosmological volume element, $dn/dM$ is the halo mass function from \cite{Tinker:2008}, and $\bar{y}_\ell$ and $\bar{\kappa}_\ell$ are the harmonic space profiles of the Compton-$y$ map and the lensing convergence, respectively. The two-halo term is given by
\begin{equation}
\label{eq:Cl2h}
C_{\ell}^{\kappa,y; 2h} = \int_{z_{\rm{min}}}^{z_{\rm{max}}} dz \frac{dV}{dz d\Omega} b_{\ell}^{\kappa}(z) \ b_{\ell}^{y}(z) \ P_{\rm{lin}}(k,z),
\end{equation}
where $k = (\ell + 1/2)/\chi$, $\chi$ is the comoving distance to redshift $z$, $P_{\rm{lin}}(k,z)$ is the linear power spectrum, and $b^{\rm \kappa}_\ell$ and $b^{\rm y}_\ell$ are the effective linear bias parameters describing the clustering of the two tracers.
% $P_{\rm lin}(k,z)$ is the linear matter power spectrum. Here we denote the 2-halo 3D power spectrum as $P^{\rm A,B; 2h}(k=(\ell + 1/2)/\chi,z) = b_{\ell}^{\rm A}(z) \ b_{\ell}^{\rm B}(z) \ P_{\rm{lin}}(k,z)$. 

The total power spectrum is obtained by summing the 1-halo and 2-halo components:
%%
%\begin{equation}\label{eq:Cltot}
%    C_{\ell}^{\rm A,B} = \int_{z_{\rm{min}}}^{z_{\rm{max}}} dz \frac{dV}{dz %d\Omega} \bigg( (P^{\rm A,B; 1h}(k,z))^{\alpha^s_{\rm AB}} +  (P^{\rm A,B; %2h}(k,z))^{\alpha^s_{\rm AB}} \bigg)^{1/{\alpha^s_{\rm AB}}},
%\end{equation}
%%
%where $\alpha^s_{\rm AB}$ is the softening parameter that impacts the profile %in 1-halo to 2-halo transition regime. Setting $\alpha^s_{\rm AB} = 1$ reduces %the above equation of total power spectra into familiar sum of 1-halo and %2-halo components:
%
\begin{equation}\label{eq:Cltot}
    C_{\ell}^{\kappa,y} = C_{\ell}^{\kappa,y; 1h} + C_{\ell}^{\kappa,y; 2h}.
\end{equation}
%
%
%The exact form of $b_{\ell}$, $\bar{u}_{\ell}$, $\alpha^s$ and the values of $M_{\rm{min}}$, $M_{\rm{max}}$ will depend on the particular fields being correlated; we will describe these quantities in the following sections. 
%

%The real space correlation functions can be obtained from inverse Fourier transform of Eq.~\ref{eq:Cltot}.  In particular, 
The $y$-$\gamma_t$ cross-correlation can then be written as (similar to \citep{Hojjati20117} but without the flat-sky approximation):
\begin{equation}
\label{eq:xigty}
\xi^{\gamma_t y}(\theta) = \int \frac{d\ell \ \ell}{2\pi} J_{2}(\ell \theta) C_{\ell}^{\kappa,y},
\end{equation}
where $J_{2}$ is the second order Bessel function of the first kind.

\subsection{The halo pressure profile}\label{sect:compton-y_theo}

\begin{figure*}
\includegraphics[width=1.\textwidth]{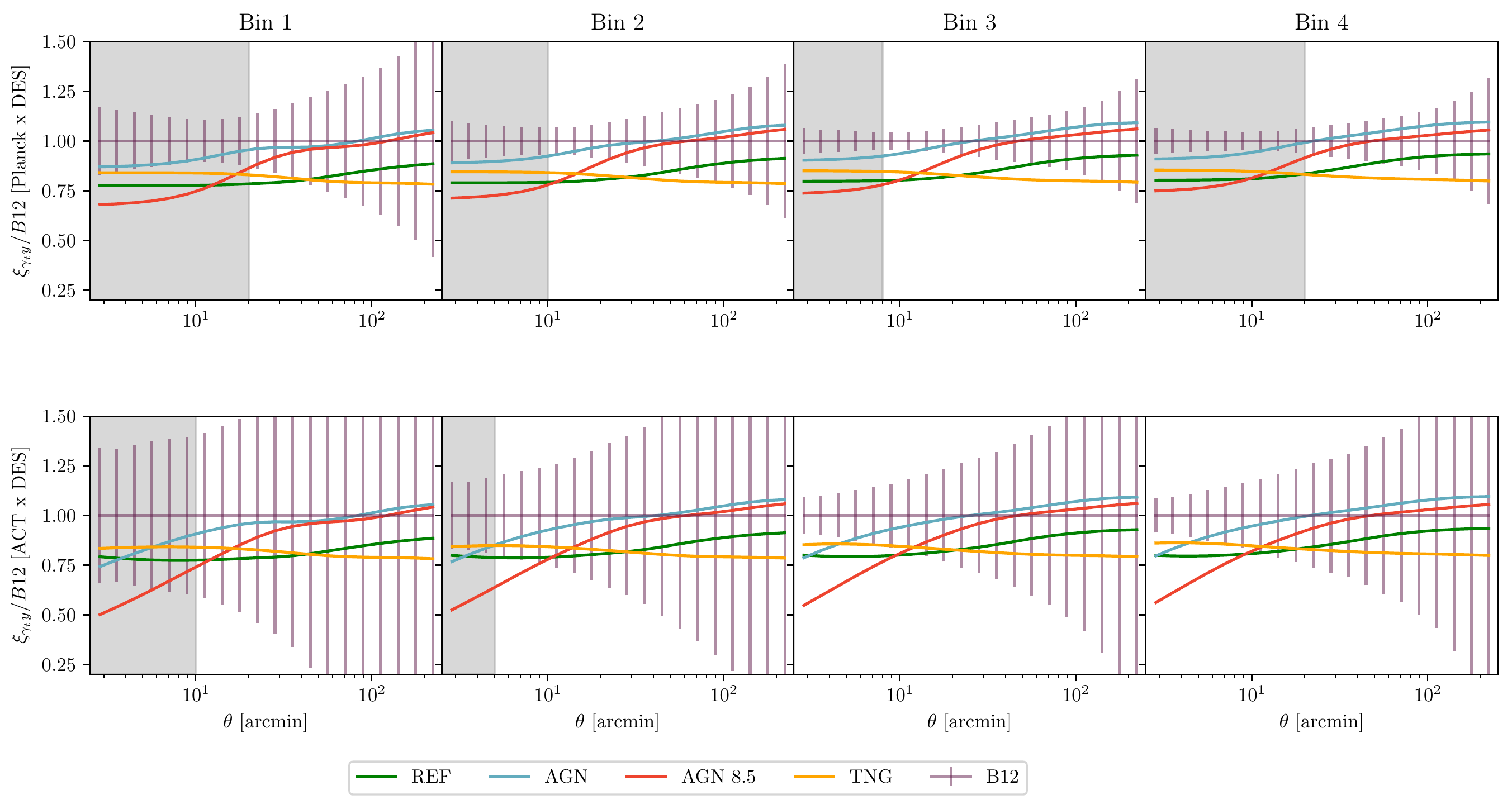}
\caption[]{Differences in the expected $\xi^{\gamma_t y}$ correlation signals assuming different pressure profile models, as introduced in \S~\ref{sect:compton-y_theo}. We take the predicted signal obtained with the B12 model as a reference. We show the predicted signals for the four tomographic bins of the DES shape catalog and for the two different Compton-$y$ maps (ACT and \textit{Planck}). For this figure, the shear part of the signal has been modelled assuming a  Navarro–Frenk–White (NFW) profile for the DM profile. Error bars for the B12 model show the expected measurement uncertainties ACT measurement uncertainties are larger than \textit{Planck} because of the smaller sky coverage. Grey shaded regions indicate the scales that are not used in this analysis (Table \ref{table_scale_cuts}). The error bars are strongly correlated between bins.}
\label{theory_difference_models}
\end{figure*}

\begin{figure}
\includegraphics[width=.45\textwidth]{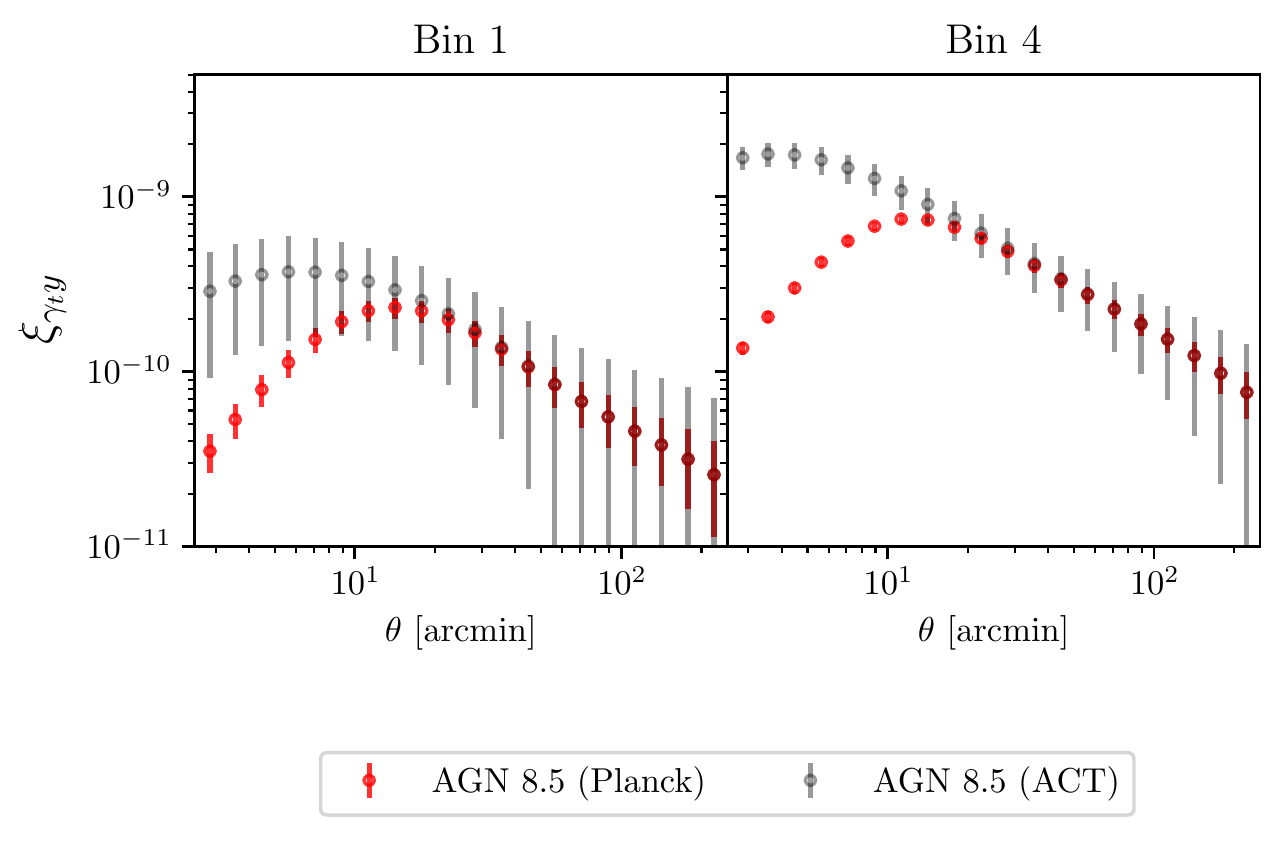}
\caption[]{Expected $\xi^{\gamma_t y}$ correlation signals for the ACT x DES (grey) and \textit{Planck} x DES (red) measurements, assuming AGN 8.5 feedback model, as introduced in \S~\ref{sect:compton-y_theo}. We just show the first and the fourth bin; error bars represent measurement uncertainties. The dampening of the signal predicted for the \textit{Planck} map at small scales is due to the large \textit{Planck} beam.}
\label{theory_models}
\end{figure}

The profile in harmonic space of the Compton-$y$ map can be {related to the } pressure profile $P_e(x|M_{200c},z)$ via (see, e.g., \cite{Komatsu1999,komatsu2002,hill2013}):
\begin{multline}\label{eq:uyl}
\bar{y}_{\ell}(M_{200c},z) = b^j(\ell) \frac{4\pi r_{200c}}{l^2_{200c}} \frac{\sigma_T}{m_e c^2} \int_{x_{\rm{min}}}^{x_{\rm{max}}} dx \ x^2 \ P_e(x|M_{200c},z) \\ \times \frac{\sin(\ell x/l_{200c})}{\ell x/l_{200c}}.
\end{multline}
In the above equation we have defined $x=a(z)R/R_{200c}$, where $a(z)$ is the scale factor, $R$ is the radius and $R_{200c}$ the radius enclosing the spherical region in which the average density is 200 times the critical density at the respective redshift; moreover, we defined  $l_{200c} = D_A/R_{200c}$, where $D_A$ is the angular diameter distance to redshift $z$. We choose $x_{\rm{min}} = 10^{-3}$ and $x_{\rm{max}} = 4$, %\footnote{We did experiment slightly changing the value of $x_{\rm{max}}$, but found that our results were basically unaffected.},
which ensures that the above integral captures the contribution to the pressure from the extended profile of hot gas. Lastly, the term $b^{j}(\ell) = \exp{[-\ell(\ell + 1) \sigma_j^2/2]}$ captures the beam profile. Here $\sigma_j = \theta_{j}^{\rm FWHM}/\sqrt{8\ln 2}$ and we have $\theta_{1}^{\rm FWHM} = 10$ arcmin and  $\theta_{2}^{\rm FWHM} = 1.6$ for \textit{Planck} and ACT respectively.{ We note that since we are dealing with pixellised Compton-$y$ maps, we should also take into account the pixel window function; in practice, since the size of the pixel is always smaller than $0.5 \theta^{\rm FWHM}$, it can be safely neglected.}

The effective tSZ bias $b_\ell^y$ is given by:
\begin{equation}
\label{eq:byl}
b_\ell^y(z) = \int_{M_{\rm{min}}}^{M_{\rm{max}}} dM \ \frac{dn}{dM} \bar{y}_\ell(M,z) b_{\rm{lin}}(M,z),
\end{equation}
where $b_{\rm{lin}}$ is the linear bias of halos with mass $M$ at redshift $z$ (in the halo model, the halos are biased tracers of the underlying linear matter field). We use the \citet{Tinker:2010} fitting function for halo bias as a function of mass and redshift.

In this work, we consider the following pressure profile models, calibrated against hydrodynamical simulations implementing different baryonic feedback prescriptions:
\begin{itemize}
    \item The \citet{Battaglia_2012} (B12) pressure profile,  calibrated against a suite of hydrodynamical TreePM-SPH simulations that include radiative cooling, star formation, supernova feedback, and AGN feedback;
    \item The \citet{LeBrun15} REF model, calibrated against a version of the cosmo-OWLs simulations \citep{OWL} that includes prescriptions for radiative cooling, stellar evolution, mass-loss, chemical enrichment and kinetic stellar feedback;
    \item The \citet{LeBrun15} AGN model,  calibrated against a version of the cosmo-OWLs simulations that also includes a prescription for AGN feedback. In particular, we include the two variants, the AGN and AGN 8.5 models, with the latter being characterised by a stronger AGN feedback prescription;
    \item The pressure profile as measured in the IllustrisTNG simulation (TNG hereafter; \citet{Springel2018}). The pressure profile is modelled as a generalised Navarro–Frenk–White (NFW) profile \citep{nfw:1996} similar to the B12 model, but fitting the model parameters (as well as their halo mass and redshift evolution) to the pressure profiles measured in the IllustrisTNG simulation.\footnote{In particular, we followed \cite{Moser2021} and measured the pressure profile parameters dividing the halos of the simulations in two halo mass bins (M$\sim 10^{13.5}-10^{14.25}$ \msun, M$\sim10^{14.25}-10^{15.0}$ \msun), and at three different redshift ($z=0,0.31,0.6$), and interpolated the mass/redshift dependence. We could not use more bins due to the paucity of halos in this mass and redshift range.}
\end{itemize}

{These are a wide range of hydrodynamical simulations with (more or less) different AGN prescriptions (see, e.g. EAGLE simulation, \cite{Hellwing2016}, Horizon simulation, \cite{Chisari2018},  BAHAMAS simulation \cite{bahamas}, etc.); measuring and comparing to all the pressure profiles from these simulations goes beyond the scope of this work. Moreover, we believe the profiles considered here already span a sufficiently wide range of different feedback models.}

It is important to note that we are not interested here in freeing the parameters of the pressure profile models developed by \citet{Battaglia_2012} and \citet{LeBrun15}, but rather we want to use their best fit values to test whether the feedback mechanisms implemented in the simulations provide a good description of our measurement (within uncertainties). A different approach, where the pressure profile parameters are varied, is adopted in \paperB.

We show how different pressure profile models translate to differences in the expected $\xi^{\gamma_t y}$ correlation signals in Fig.~\ref{theory_difference_models} (see also Fig.~\ref{theory_models} for the expected amplitude of the signal). Different predictions are obtained via Eq. \ref{eq:xigty}, assuming fiducial values for all the ingredients of the modelling except for the pressure profiles. For this comparison, the shear part of the signal has been modelled assuming a NFW profile for the DM profile (we note, however, that when analysing our data we would also allow the DM profile to vary under the effects of baryons, as explained in the next section). We show the $\xi^{\gamma_t y}$  correlation signals for the \textit{Planck} and ACT Compton-$y$ maps with the four tomographic bins of the DES shape catalog. The modelling of the \textit{Planck} x DES and ACT x DES measurements differ only in the amplitude of the FWHM of the beam. We take the B12 model as a reference. The REF model is characterised by a $10-20$\% lower amplitude at all angular scales and for all the different redshifts. In this model, a large fraction of halo baryons are able to cool and form stars, reducing the gas fraction and the tSZ amplitude. On the other hand, the two AGN models show a similar amplitude to the B12 model at large scales, but the most extreme AGN model (8.5) shows a significant lower amplitude (down to $\sim 40$\% in the ACT x DES measurement) at small scales, related to the gas ejection from the halo due to AGN feedback. Lastly, the TNG AGN model, which is based on a different suite of simulations and different AGN feedback mechanism compared to all the other models, is characterised by a 20\% lower amplitude at all scales. The TNG AGN feedback is neither able to heat up the gas as much as the B12 model, resulting in a lower amplitude at all scales, nor to eject the gas from the halo as efficiently as the most extreme AGN 8.5 scenario (which would cause a lower amplitude at small scales).

%heat the gas in the outer regions

\subsection{Shear signal}\label{sect:shear_signsl}
The lensing part of our signal is described by the profile of the lensing convergence in harmonic space:
\begin{equation}\label{eq:ukl}
    \bar{\kappa}_{\ell}(M_{\rm vir},z) = \frac{W^\kappa(z)}{\chi^2}  u_{\rm{m}}(k,M_{\rm{vir}}),
\end{equation}
where $k = (\ell + 1/2) / \chi$ and $u_m(k,M)$ is the Fourier transformation of the dark matter density profile.  The quantity $W^\kappa(z(\chi))$ is the lensing kernel, given by:
\begin{equation}
\label{eq:lensing_kernel}
    W^\kappa(z(\chi)) = \frac{3 H_0^2 \Omega_m}{2 c^2} \frac{\chi}{a(\chi)} \int_{\chi}^{\infty} d\chi' n_{\kappa}(z(\chi')) \frac{dz}{d\chi'}\frac{\chi' - \chi}{\chi'},
\end{equation}
with $n_{\kappa}$ the normalized redshift distribution of the source galaxies. The redshift distribution of the source galaxies peaks at significantly higher redshift compared to the sensitivity of our signal (see Fig. 2 of \paperB); this implies that the dilution of the signal due to sources physically associated to foreground clusters \citep{Sheldon2004,y3-gglensing} is negligible. The effective lensing bias is:
\begin{equation}
\label{eq:bkl}
b_\ell^{\kappa}(z) = \int_{M_{\rm{min}}}^{M_{\rm{max}}} dM \ \frac{dn}{dM} \bar{\kappa}_\ell(M,z) b_{\rm{lin}}(M,z),
\end{equation}
where $b_{\rm{lin}}$ is the linear bias of halos with mass $M$ at redshift $z$ which we model using the \citet{Tinker:2010} fitting function. In case of no feedback, the dark matter profile can be modelled by a simple NFW profile \citep{nfw:1996}, but in practice baryonic feedback can affect the overall matter distribution and matter profile. To model this effect we take two approaches.

In a first approach, we simply re-scale the lensing profile by a mass-independent factor that reads:
\begin{equation}
\bar{\kappa}_\ell(M,z) \rightarrow \bar{\kappa}_\ell(M,z) \sqrt{\frac{P_{\rm DM + baryons} (k,z)}{P_{\rm DM}(k,z)}},
\end{equation}
where $k = (\ell + 1/2)/\chi$, $P_{\rm DM}$ and $P_{\rm DM + baryons}$ are the power spectrum from a dark-matter only simulation and the power spectrum from a hydrodynamical simulation with dark-matter and a sub-grid prescription for baryonic effects. This approach is equivalent to the one assumed in some cosmic shear analyses, where the effect of baryonic feedback processes is taken into account by re-scaling the 3D matter power spectrum \citep{Troxel2018,Hikage2019}.  When testing the REF, AGN and AGN 8.5 models, we re-scaled the lensing profile using the power spectra measured directly in the corresponding cosmo-OWLs simulations (as reported by \citet{vanDaalen2020}). The effect of the re-scaling for the REF model is below 1\% at all scales, whereas for the AGN and AGN 8.5 models it mostly dampens the amplitude of $\xi^{\gamma_t y}$ below 10 arcminutes, reaching a 10\% dampening at 2.5 arcminutes for the AGN 8.5 model (which is the most affected model). %The effect of this re-scaling is smaller than the choice of pressure profile has on $\xi^{\gamma_t y}$, as shown by Fig.~\ref{theory_difference_models}. 
For the B12 model we do not have at our disposal the 3D power spectra measured in the corresponding simulations with and without baryonic feedback; so we did not consider this model in this first approach.

%we arbitrarily re-scaled the lensing kernel using the same power spectra used for the AGN model, but we will quote the corresponding result in parenthesis to remind the reader about this caveat.

In the second approach we model the effects of baryonic feedback on the lensing kernel with more flexibility. Instead of using a re-scaled version of the NFW profile, we use the Mead model \citep{Mead2015} to determine $u_m(k,M)$, the Fourier transformation of the dark matter density profile. The Mead model builds upon the NFW profile, but it adds additional freedom such that the model can capture the effect of baryonic physics on the internal structure of halos. This is achieved by adding two parameters: $A_{\rm Mead}$, the amplitude of the concentration-mass relation, and $\eta_{\rm Mead}$, the ``halo bloating parameter'', which produces a (mass dependent) expansion of the halo profile.\footnote{{We note that in \cite{Mead2015}, the authors provide best-fit values for the parameters $A_{\rm Mead}$ and $\eta_{\rm Mead}$ for a number of hydrodynamical simulations, as well as suggesting a relation between the two. We cannot use those values or such a relation here, as our implementation of the Mead model is slightly different from the one presented in \cite{Mead2015}, which is \textit{optimised} for a cosmic shear analysis}. } The NFW profile is still included in the Mead model parameter space, as well as the re-scaled versions of the NFW profile used in the aforementioned approach. Full expressions for $u_m(k,M)$ and for the effective linear bias parameter $b^{\kappa}_{\ell} (z)$ of the Mead model are provided in \paperB. In this second approach, when testing different feedback models, we marginalise over $A_{\rm Mead}$ and $\eta_{\rm Mead}$  using wide priors (similar to the ones assumed by \cite{Maccrann2017}). This is more conservative than re-scaling the NFW profile using the power spectra measured in hydrodynamical simulations. Indeed, in the first approach we assumed the re-scaling of the lensing kernel to be independent of halo mass; our measurement, however, is mostly sensitive to $M_{200c} \sim 10^{14}$ \msun (see \paperB), so if the effect of baryonic feedback models were halo mass dependent, the re-scaling might be not accurate. By marginalising over the Mead model parameters, we let the data re-scale the lensing profile by the ``right'' amount. Note that \textit{a priori} there should be a relation between the Mead halo model parameters and the pressure profile parameters, as we expect baryonic processes to have a simultaneous impact on the matter and gas. As the Mead halo model implemented in this work is an heuristic model, it is hard to place physically motivated priors on such a relation; therefore, we consider the Mead model and the pressure profile parameters as independent. In this respect, a more coherent frameworks (e.g., \cite{Mead2020b}) where the shear and tSZ signals are modelled starting from the distribution of gas, matter and stars can provide better insights into the interplay between the pressure and matter profiles in the presence of baryonic feedback processes.

We provide in Appendix~\ref{sect:sims_buzzard} further validation of our modelling by measuring the shear-Compton-$y$ map cross correlation on the fiducial DES Y3 N-body simulations.

\subsection{Astrophysical and nuisance parameters}

\begin{table}
%\tiny
\caption {Cosmological, systematic and astrophysical parameters. The cosmological parameters considered are $\Omega_{\rm m}$, $\sigma_8$, $\Omega_{\rm b}$ (the baryonic density in units of the critical density),  $n_s$ (the spectral index of primordial density fluctuations) and $h$ (the dimensionless Hubble parameter). The nuisance parameters are the multiplicative shear biases $m_i$ and the photometric uncertainties in the mean of the weak lensing samples $\Delta z_i$. The astrophysical parameters $A_{\rm IA,0}$ and $\alpha_{\rm IA}$ describe the intrinsic alignment model, whereas {$A_{\mathrm {MEAD}}$} and {$\eta_{\mathrm {MEAD}}$} are the Mead halo model parameters. The column ``range'' indicates the parameters space spanned when sampling the parameters posterior during the analysis. We report the boundaries for both Flat and Gaussian priors. For Gaussian priors we also report the mean and the $\sigma$ values in the prior column. {Priors are described in \S~\ref{sect:like}.}}
\centering
%\begin{adjustbox}{width=0.7\textwidth}
\begin{tabular}{|c|c|c|}
 \hline
\textbf{Parameter} & \textbf{Range}& \textbf{Prior}\\
 \hline
{$\Omega_{\rm m}$ [\textit{Planck}]} &$0.2 ... 0.4$& 0.315 $\pm$  0.007 \\
{$\sigma_{\rm 8}$ [\textit{Planck}]} &$0.6 ... 1.1$& 0.811 $\pm$  0.006 \\
{$\Omega_{\rm m}$ [{DES}]} &$0.2 ... 0.4$& 0.27 $\pm$ 0.02\\
{$\sigma_{\rm 8}$ [{DES }]} &$0.6 ... 1.1$& 0.82 $\pm$ 0.05\\
 \hline
{$h$} & {\rm Fixed} &  0.674\\ 
{$n_{\rm s}$} & {\rm Fixed}& 0.965\\ 
{$\Omega_{\rm b}$} & {\rm Fixed}& 0.0493 \\
 \hline
{$\Delta m_1\times10^2$} &$-10.0 ... 10.0$& -0.63 $\pm$ 0.91 \\
{$\Delta m_2\times10^2$} &$-10.0 ... 10.0$& -1.98 $\pm$ 0.78 \\
{$\Delta m_3\times10^2$} &$-10.0 ... 10.0$& -2.41 $\pm$ 0.76 \\
{$\Delta m_4\times10^2$} &$-10.0 ... 10.0$& -3.69 $\pm$ 0.76 \\

{$\Delta z_1\times10^2$} &$-10.0 ... 10.0$& 0.0 $\pm$ 1.8 \\
{$\Delta z_2\times10^2$} &$-10.0 ... 10.0$& 0.0 $\pm$ 1.5 \\
{$\Delta z_3\times10^2$} &$-10.0 ... 10.0$& 0.0 $\pm$ 1.1 \\
{$\Delta z_4\times10^2$} &$-10.0 ... 10.0$& 0.0 $\pm$ 1.7 \\

 \hline
{$A_{\mathrm {IA},0}$} &$-5.0 ... 5.0$& 0.49 $\pm$ 0.15 \\
{$\eta{\mathrm {IA}}$} &$-5.0 ... 5.0$& Flat \\
 \hline
{$A_{\mathrm {MEAD}}$} &$-5.0 ... 5.0$& Flat \\
{$\eta_{\mathrm {MEAD}}$} &$0 ... 1.0$& Flat \\
 \hline
\end{tabular}
%\end{adjustbox}
\label{table1}
\end{table}
Astrophysical and measurement systematic effects are modelled through nuisance parameters. When performing our analysis, we marginalise over all the nuisance parameters. Values and priors are summarised in Table~\ref{table1}.

\textit{Photometric redshift uncertainties}. Uncertainties in the photometric redshift estimates from the SOMPZ method for the 4 tomographic bins of the weak lensing sample are parametrised through a shift $\Delta z$ in the mean of the redshift distributions:
\begin{equation}
n^i(z) = \hat{n}^i(z-\Delta z ),
\end{equation}
where $\hat{n}^i$ is the original estimate of the redshift distribution coming from the photometric redshift code for the bin $i$. This parameterisation of the redshift uncertainties has shown to be adequate for the DES Y3 analysis \cite{y3-hyperrank}. We assume DES Y3 Gaussian priors for the shift parameters.

\textit{Multiplicative shear biases}. {Biases coming from the shear measurement pipeline are modelled through an average multiplicative parameter $1+m^i$ for each tomographic bin $i$}, which affects our measurement as $\xi^{\gamma_t^i y} \rightarrow (1+m^i)\xi^{\gamma_t^i y}$. Gaussian priors are assumed for each of the $m^i$, and have been estimated in \cite{y3-imagesims}. The major contribution to the shear multiplicative bias comes from blending effects due to source crowding.

\textit{Intrinsic galaxy alignments {(IA)}}. IA has been neglected in all the previous works on shear-Compton-$y$ cross-correlations. In theory, an IA contribution is expected, as Compton-$y$ maps trace the underlying dark matter density field.  Our implementation of the IA model follows the non-linear alignment (NLA) model \citep{Hirata2004,Bridle2007}. It can be easily incorporated in the modelling by modifying the lensing kernel (Eq. \ref{eq:lensing_kernel}):

\begin{equation}
\label{eq:ia_nla}
W^{\kappa}(\chi) \rightarrow W^{\kappa}(\chi) - A(z(\chi)) {n_{\kappa}(z(\chi))}\frac{dz}{d\chi}.
\end{equation}

The amplitude of the IA contribution can be written as a power-law:

\begin{equation}
A(z) = A_{\rm IA,0} \left(\frac{1+z}{1+z_0} \right)^{\eta_{\rm IA}} \frac{c_1 \rho_{m,0}}{D(z)},
\end{equation}
with $z_0= 0.62$, $c_1\rho_{m,0}=0.0134$ (\cite{Bridle2007}, \cite*{methodpaper}) and $D(z)$ the linear growth factor. We marginalise over $A_{\rm IA,0}$ and $\eta_{\rm IA}$ assuming the constraints from the DES Y1 3x2 analysis. More details about the IA model, and its relative strength compared to the Compton-$y$-shear signal, are given in \paperB.

We also tested an additional 1-halo IA contribution due to satellite galaxies alignment, following \cite{Fortuna2020}. This extra contribution requires modelling the Halo Occupation Distribution (HOD) of satellite galaxies and the fraction of satellites as a function of redshift, which are somewhat uncertain for the DES Y3 weak lensing sample. For this reason, we decided to remove scales where this extra 1-halo model could provide a significant contribution. More details are given in \paperB.

%and  provide an amplitude for the radial alignment signal.

\subsection{Likelihood and Covariance}\label{sect:like}

\begin{table}
%\tiny
\caption {Angular scales considered for the analysis (in arcmin). More details are provided in \S~\ref{sect:like}.}
\centering
%\begin{adjustbox}{width=0.7\textwidth}
\begin{tabular}{|c|c|c|c|c|}
 \hline

 & \textbf{bin 1} & \textbf{bin 2} & \textbf{bin 3} & \textbf{bin 4} \\

 \hline
\textit{Planck} x DES  & 20'-250'  &  8'-250'  & 8-250  &20'-250'    \\
ACT x DES  & 10'-250'  &  5'-250'  & 2.5'-250'  &2.5'-250'    \\
\hline
\end{tabular}

%\end{adjustbox}
\label{table_scale_cuts}
\end{table}

Our data vector includes shear-Compton y-map correlations $\xi^{\gamma_t y}$ from both \textit{Planck} and ACT maps. In both cases, we cross-correlated the maps with the four DES Y3 tomographic bins, for a total of 8 correlation measurements. We measured the cross-correlations in 20 bins (equally spaced in logarithmic scale) between 2.5 and 250 arcmin. We exclude some angular scales from our analysis, based on three different criteria: (1) we exclude all the scales below 8 arcmin for the \textit{Planck} measurement, since these scales are well below the \textit{Planck} beam size; (2) we remove small scales where our IA modelling might be inadequate (see \paperB); (3) we remove scales that might be significantly affected by cosmic infrared background (CIB) contamination (see \S~\ref{sect:map_tests}). Depending on the tomographic bin, we consider the most stringent criteria among these three. We summarise the scale cuts in Table~\ref{table_scale_cuts}. Having defined our data vector, in order to test our models, we evaluate the posterior of the parameters conditional on the data by assuming a multivariate Gaussian likelihood for the data, i.e.

\begin{linenomath*}
\begin{equation}
    \ln \mathcal{L}(\mathbfcal{D}|\Theta) = -\frac{1}{2} (\vec{\mathbfcal{D}} - \vec{\mathbfcal{T}}(\Theta)) \, {\varmathbb{C}}^{-1} \,  (\vec{\mathbfcal{D}} - \vec{\mathbfcal{T}}(\Theta))^{\rm T}.
\end{equation}
\end{linenomath*}
Here, $\vec{\mathbfcal{D}}$ is the measured $\xigty$ data vector of length $N_{\rm data}$, $\vec{\mathbfcal{T}}$ is the theoretical model for the data vector at the parameter values given by  $\Theta$, and ${\varmathbb{C}}^{-1}$ is the inverse covariance matrix of shape $N_{\rm data} \times N_{\rm data}$.  The measurement covariance is {modelled from theory, including both a Gaussian and a non Gaussian term;} we provide a detailed description in Appendix~\ref{sec:covariance}, along with validation based on data and simulations. The posterior is then the product of the likelihood and the priors:
\begin{linenomath*}
\begin{equation}
    \mathcal{P}(\Theta | \mathbfcal{D}) = \frac{\mathcal{L}(\mathbfcal{D}|\Theta) {\rm P}(\Theta)}{{\rm P}(\mathbfcal{D})}
\end{equation}
\end{linenomath*}
where the ${\rm P}(\Theta)$ are the priors on the parameters of our model, and ${\rm P}(\mathbfcal{D})$ is the evidence of data. To sample the posteriors of our parameters, we use the \textsc{Polychord} sampler \citep{poly1,poly2}, which is a nested sampler that uses slice sampling to sample within the nested iso-likelihood contours.
 
 %$\Omega_{\rm m}$ (the matter density in units of the critical density), $\Omega_{\rm b}$ (the baryonic density in units of the critical density), $\sigma_8$ (the amplitude of structure in the present day Universe, parameterised as the standard deviation of the linear overdensity fluctuations on a 8$h^{ - 1}$ Mpc scale), $n_s$ (the spectral index of primordial density fluctuations) and $h$ (the dimensionless Hubble parameter). 

For the cosmological parameters, we assume a flat $\Lambda$CDM cosmology and vary two parameters:  $\Omega_{\rm m}$ and $\sigma_8$, leaving $\Omega_{\rm b}$ (the baryonic density in units of the critical density),  $n_s$ (the spectral index of primordial density fluctuations) and $h$ (the dimensionless Hubble parameter) fixed to values from \citet{Planck}. For $\Omega_{\rm m}$ and $\sigma_8$, and for each model under testing we run our analysis twice, with Gaussian priors centered in \cite{Planck} and DES Y1 \cite{desy1_3x2} values, with width equal to the 1$\sigma$ uncertainty on the two parameters from the two analyses, respectively. We used DES Y1 priors as at the time of writing the DES Y3 analysis had not been released yet. We also  marginalise over nuisance parameters describing photo-$z$ uncertainties, shear biases and IA effects in our measurements. We use fiducial DES Y3 priors for the photo-$z$ uncertainties and shear biases. As for the IA priors, we use a Gaussian prior on the amplitude of the signal as constrained by the DES Y1 3x2 analysis \citep{Samuroff2019}. We also marginalise over the Mead halo model parameters {$A_{\mathrm {Mead}}$} and {$\eta_{\mathrm {Mead}}$} using broad flat priors.

All the parameters varied in this work and priors assumed are given in Table \ref{table1}. We discuss in detail the impact of the priors on our measurement in \S~\ref{sect:feedback_models} (see Fig.~\ref{priors_models}).

\section{Results}
We first present our measurement with the relevant systematic tests, and then we show the comparison with the theoretical predictions using the halo model framework and pressure profiles from hydrodynamical simulations.

\subsection{TSZ-weak lensing cross-correlation estimator}
Our estimator for the tSZ-weak lensing cross-correlation is the cross-correlation $\xi^{\gamma_t y}$, which is constructed starting from the DES catalog of galaxy shapes and from the ACT and \textit{Planck} Compton-$y$ maps.  Concerning the Compton-$y$ maps, we create a catalog of points using the coordinates of the centre of each \texttt{HEALPY} pixel of the maps. For each point we then consider the complex shear of a given galaxy in the DES catalog at an angular separation $\theta$. We then compute the tangential component of the shear and multiply it by the Compton-$y$ value. We average over all the galaxies at that angular separation and over all the points in the Compton-$y$ maps. %We choose to measure the cross-correlations in 20 bins (equally spaced in logarithmic scale) between 2.5 and 250 arcmin; in the case of the measurement involving \textit{Planck} data, however, we exclude data points at $\theta < 8$ arcmin, as these are well below \textit{Planck} beam FWHM. 
The DES catalog is divided into four tomographic bins, hence we measure 8 correlation functions in total. For the computation of the signal, we use \texttt{TreeCorr} \citep{Jarvis2004}.

\subsection{Fiducial measurement}\label{sect:measurement}

%\SP{A subsection on estimator and treecorr would be good here I guess}

\begin{figure*}
\includegraphics[width=1.\textwidth]{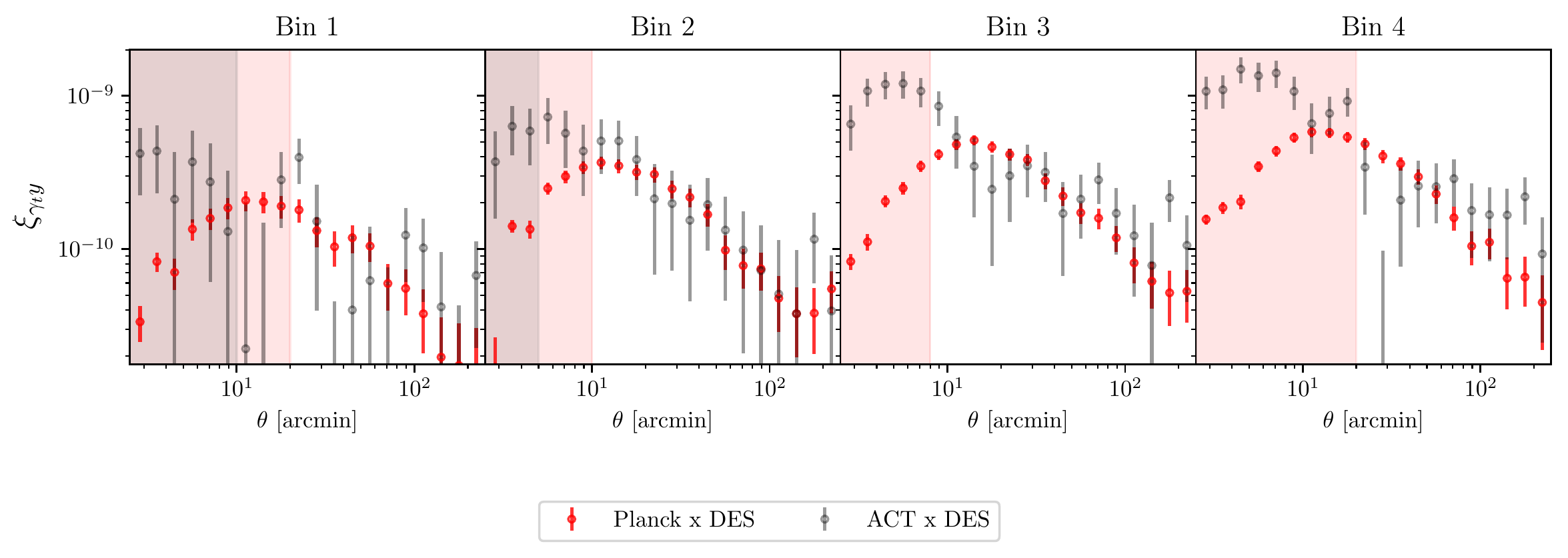}
\caption[]{Measured shear-Compton-$y$ cross-correlation $\xi^{\gamma_t y}$ from data (\textit{Planck} x DES in red ACT x DES in grey). Error bars are taken to be the square-root of the diagonal of the theory covariance presented in Appendix~\ref{sec:covariance}. The edges of the tomographic bins are [0.0, 0.358, 0.631, 0.872, 2.0], although we note that the redshift distributions are not sharply bounded (see \S~\ref{sect:desdata}). We note that the difference at small scales between the \textit{Planck} and ACT measurement is driven by the different beam size (the \textit{Planck} map has a resolution of 10 arcminutes, whereas the ACT map has a resolution of 1.6 arcminutes). The effect of the \textit{Planck} beam becomes negligible above 20 arcminutes, with the two measurements indeed showing a good agreement at large scales. Red shaded and grey shaded regions represent the scales removed from the main analysis, for the \textit{Planck} x DES and ACT x DES measurements, respectively. }
\label{measurement}
\end{figure*}

%\begin{figure*}
%\includegraphics[width=1.\textwidth]{./figs/planck_ACT_compariso%n.pdf}
%\caption[]{Comparison between the measured shear-Compton-$y$ %cross-correlation $\xi^{\gamma_t y}$ using the ACT map and the %\textit{Planck} map restricted to the ACT area. We also show as %a green band the fiducial measurement using the portion of the %\textit{Planck} map that does not overlap with the ACT %footprint.}
%\label{measurement_planck_act_area}
%\end{figure*}

\begin{table}
%\tiny
\caption {Measurements signal-to-noise ratio (SNR), defined as ${\rm SNR=\sqrt{\chi_{\rm null}-d.o.f.}}$, with $\chi_{\rm null} = \xi^{\gamma_t y}C^{-1}\xi^{\gamma_t y}$. When computing the SNR we exclude scales $\theta<8$ arcmin for the \textit{Planck} measurement, as these scales are significantly smaller than the \textit{Planck} beam. }
\centering
%\begin{adjustbox}{width=0.7\textwidth}
\begin{tabular}{|c|c|c|c|}
 \hline
\textbf{bin} & \textbf{\textit{Planck} x DES}& \textbf{ACT x DES} & \textbf{combined}\\
 \hline
bin 1 & 6.9 & 3.8 & 7.9\\
bin 2 & 11 & 3.2 & 12\\
bin 3 & 15 & 6.2 & 16\\
bin 4 & 15 & 7.5 & 17\\
 \hline
all bins & 19 & 9.3 & 21\\
\hline
\end{tabular}
%\end{adjustbox}
\label{significance}
\end{table}

%bin 1 & 6.9 & 3.8 & 7.9\\
%bin 2 & 11.2 & 3.2 & 11.6\\
%bin 3 & 14.5 & 6.2 & 15.7\\
%bin 4 & 15.1 & 7.5 & 16.8\\
% \hline
%all bins & 18.6 & 9.3 & 20.8\\

Our fiducial measurement, obtained by cross-correlating both the \textit{Planck} and ACT Compton-$y$ maps with the DES shape catalog, is presented in Fig.~\ref{measurement}. For the \textit{Planck} map, we used a version of the map where CIB contribution is de-projected (we provide more details in \S~\ref{sect:map_tests}). From Fig.~\ref{theory_models} it can be seen how the \textit{Planck} beam suppresses all the small-scale information, which is retained in the ACT x DES measurement. At large scales the two measurements are consistent with each other, although the ACT x DES measurement is noisier. Since we removed the area from the \textit{Planck} data of the DES footprint covered by ACT, the two datasets can be considered {approximately} independent. If we select only the large scales unaffected by the beams (e.g., $\theta>20$ arcmin), the two measurements are consistent with each other with a p-value = 0.05 ($\chi^2 = 60$ for 44 $d.o.f.$).\footnote{We measured the p-value by computing the difference between the two signals and used a theory covariance to estimate the covariance for the measurements difference.}  As an additional check, we also repeated the measurement using the part of the \textit{Planck} map in the ACT footprint, finding good agreement with the ACT measurement over the same area.  %Fig.~\ref{measurement_planck_act_area}, also showing a good agreement with the ACT measurement over the same area.

We report in Table~\ref{significance} the statistical significance of our measurements. Despite the higher resolution and better small-scale constraints, the ACT x DES measurement delivers lower signal-to-noise ratio (SNR) than \textit{Planck} x DES, due to the smaller sky coverage. The overall SNR for the combined measurement, considering the 4 tomographic bins together, is nonetheless improved with respect to the \textit{Planck} only measurement, and it is equal to SNR  = 21.  Concerning the individual tomographic bins, the highest SNR is provided by the correlation with the two tomographic bins with the highest redshift. The effective redshift interval probed by these two correlations is $z \sim 0.3-0.5$ (see \paperB). As a comparison, the first measurement of a cross-correlation between a CFHT convergence map and the \textit{Planck} Compton-$y$ map was detected at the $\sim 6 \sigma$ C.L. \citep{vW2014}. A stronger detection was achieved by \cite{Hojjati20117}, who detected a cross-correlation signal at the $\sim 8.1 \sigma$ C.L. for  $\xi^{\gamma_t y}$ using \textit{Planck}  data and a shape catalog from the RCSLenS survey. Our measurement improves on this, owing to its larger sky coverage of the weak lensing data.

\subsection{CIB contamination and systematic tests}\label{sect:map_tests}

% ACT-planck 0.05 (χ2=60 for 44d.o.f.)
% planck cib

\begin{table}
%\tiny
\caption {Summary of the compatibility and systematic tests performed on our measurements. We report the $\chi^2$ and the $d.o.f.$ for the difference between the \textit{Planck} and ACT measurements, as well as the difference between the measurements obtained using CIB de-projected maps and without de-projection. Lastly, for the $\gamma_{\times}$ test we report the $\chi^2$ of the null hypothesis (i.e., no signal). The number of $d.o.f$ varies depending on whether the covariance is a theory covariance (which is the case of \textit{Planck}-ACT compatibility), or a jackknife covariance (all the other cases), since for the latter case only the scales where the covariances is reliable have been used.}
\centering
%\begin{adjustbox}{width=0.7\textwidth}
\begin{tabular}{|c|c|}
 \hline
\textbf{Test} & $\chi^2/d.o.f.$\\
\hline
\textit{Planck}-ACT compatibility, all bins & 60/44\\
\hline
\textit{Planck}, CIB, bin 1 & 4/8\\
\textit{Planck}, CIB, bin 2 & 12/8\\
\textit{Planck}, CIB, bin 3 & 55/8\\
\textit{Planck}, CIB, bin 4 & 110/8\\
ACT, CIB, bin 1 & 4/8\\
ACT, CIB, bin 2 & 8/8\\
ACT, CIB, bin 3 & 14/8\\
ACT, CIB, bin 4 & 9/8\\
\hline
\textit{Planck} radio contamination  & 34/32\\
\hline
\textit{Planck}, $\gamma_{\times}$, all bins &  84/68\\
ACT,  $\gamma_{\times}$, all bins & 93/80\\
\hline
\end{tabular}
%\end{adjustbox}
\label{tests}
\end{table}

\begin{figure*}
\includegraphics[width=1.\textwidth]{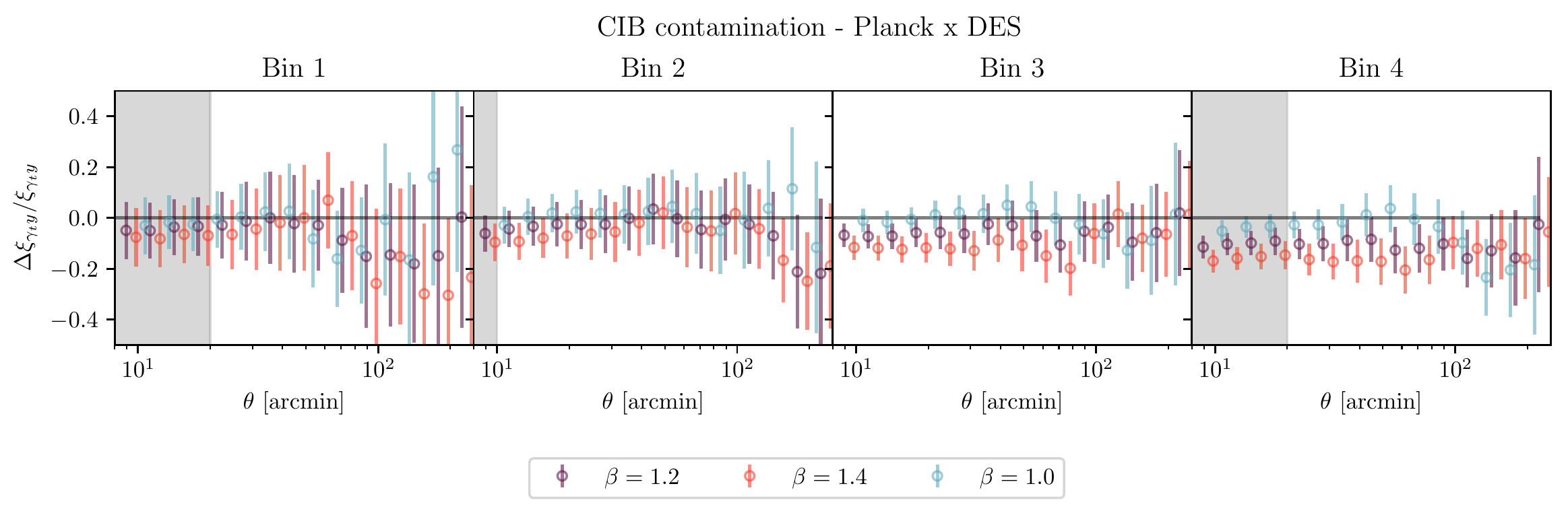}
\includegraphics[width=1.\textwidth]{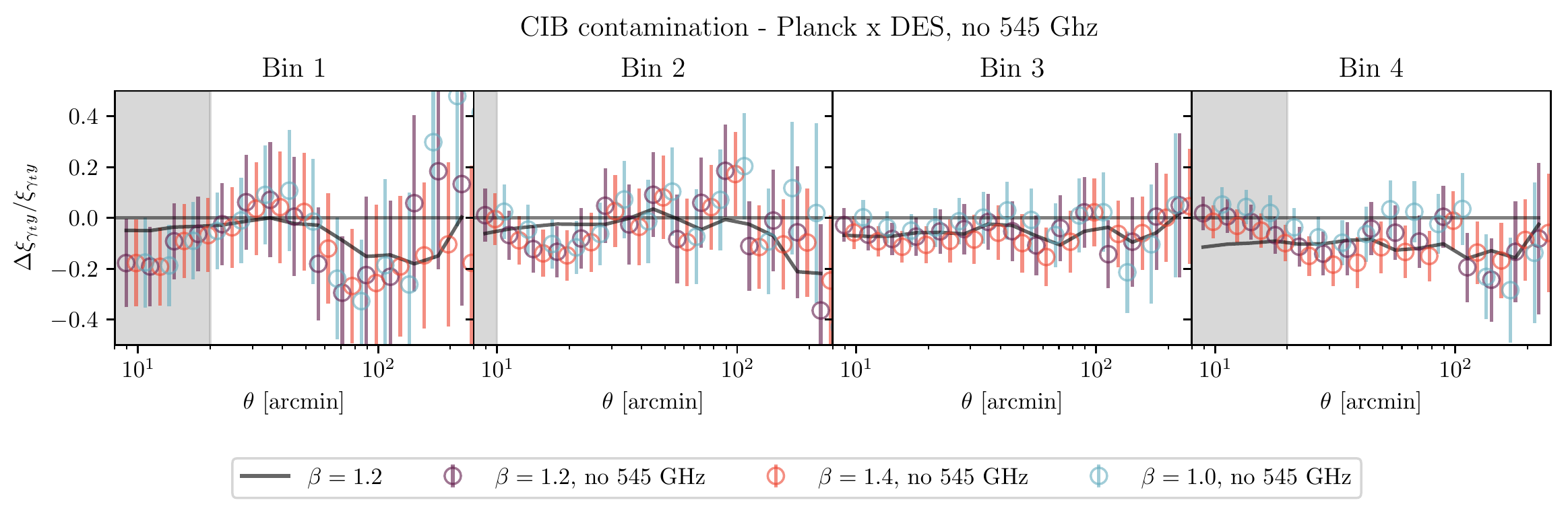}
\includegraphics[width=1.\textwidth]{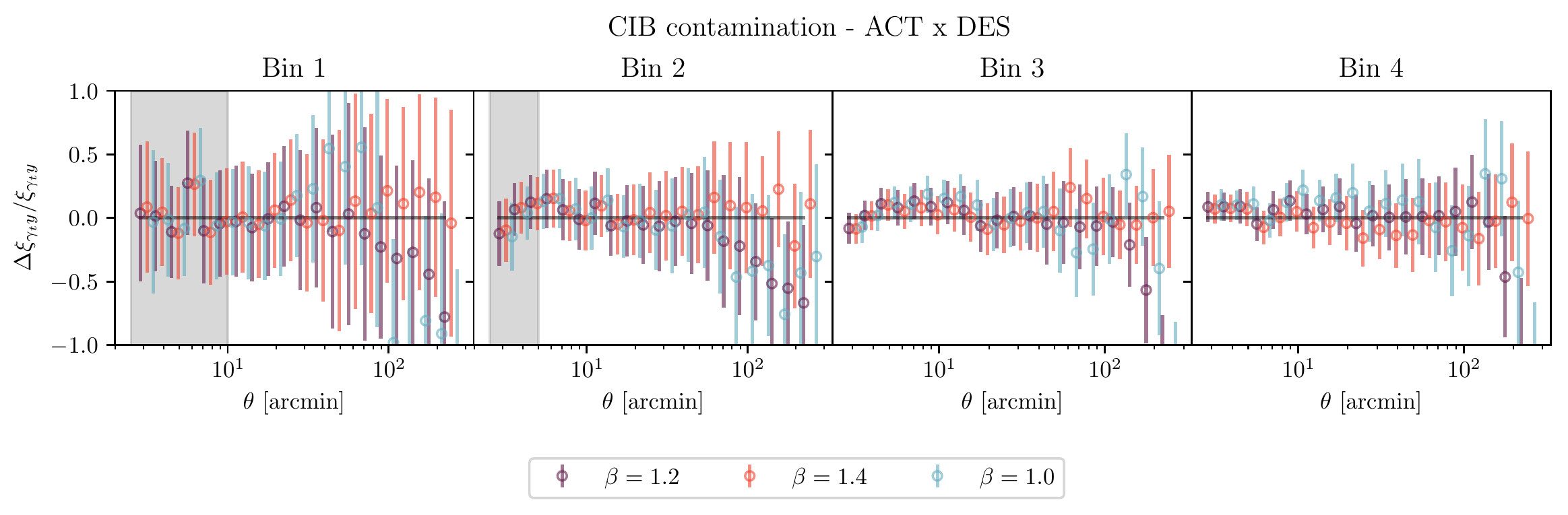}
\caption[]{Fractional difference in the measured shear-Compton-$y$ cross-correlation $\xi^{\gamma_t y}$ when computed using the Compton-$y$ map after de-projecting the CIB component, with respect to the map with no CIB de-projection. The upper plots show the results for the \textit{Planck} maps created using all the frequency channels between 30 GHz and 545 GHz; the central plots show the results for the \textit{Planck} maps excluding the 545 GHz frequency channel. In the central panels, the solid black line shows the $\beta=1.2$ measurement when including the 545 GHz channel (from the upper panel plots) as a comparison. CIB has a strong impact on the \textit{Planck} x DES measurements, especially for the ones involving the two highest redshift bins (see upper panel). CIB contamination is also seen when removing the 545 GHz frequency channel (central panels). The lower panels show the results for the ACT maps, which indicates negligible CIB contamination. Grey shaded regions represent the scales removed from the main analysis. The data are strongly correlated.}
\label{measurement_CIB}
\end{figure*}
\begin{figure*}
\includegraphics[width=1.\textwidth]{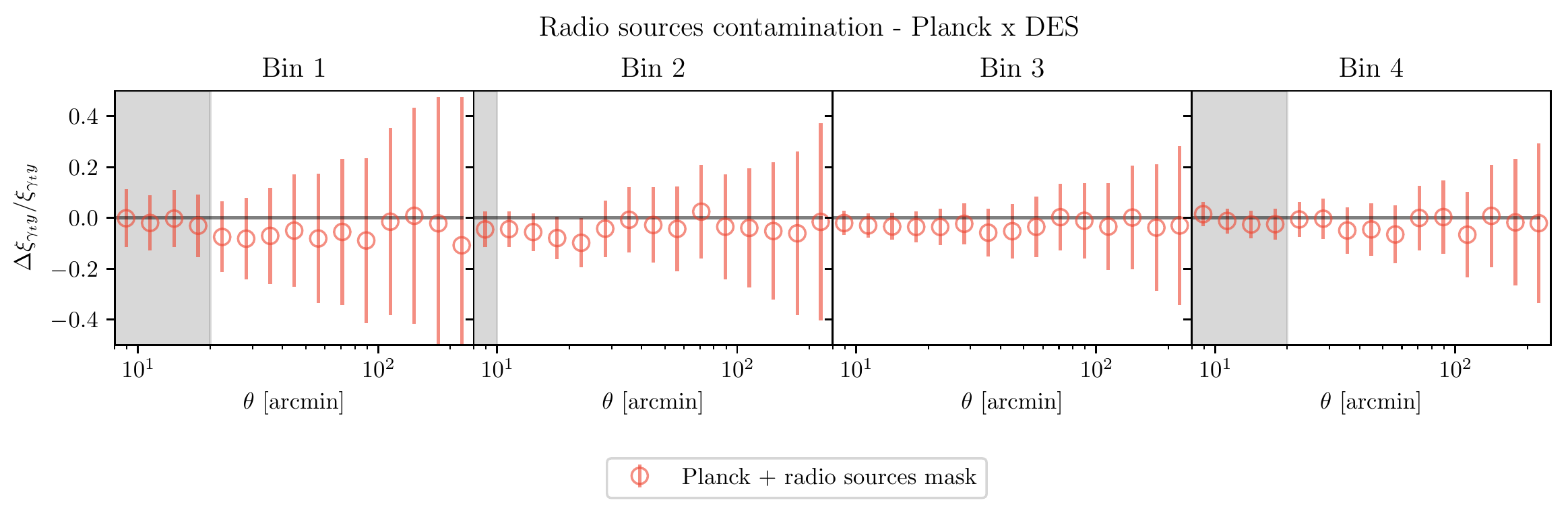}
\caption[]{Fractional difference in the measured shear-Compton-$y$ cross-correlation $\xi^{\gamma_t y}$ when computed using the \textit{Planck} Compton-$y$ map with and without masking radio sources. No statistically significant difference is measured, indicating no contamination due to radio sources. Grey shaded regions represent the scales removed in the main analysis.}
\label{measurement_planck_radio}
\end{figure*}
\begin{figure*}
\includegraphics[width=1.\textwidth]{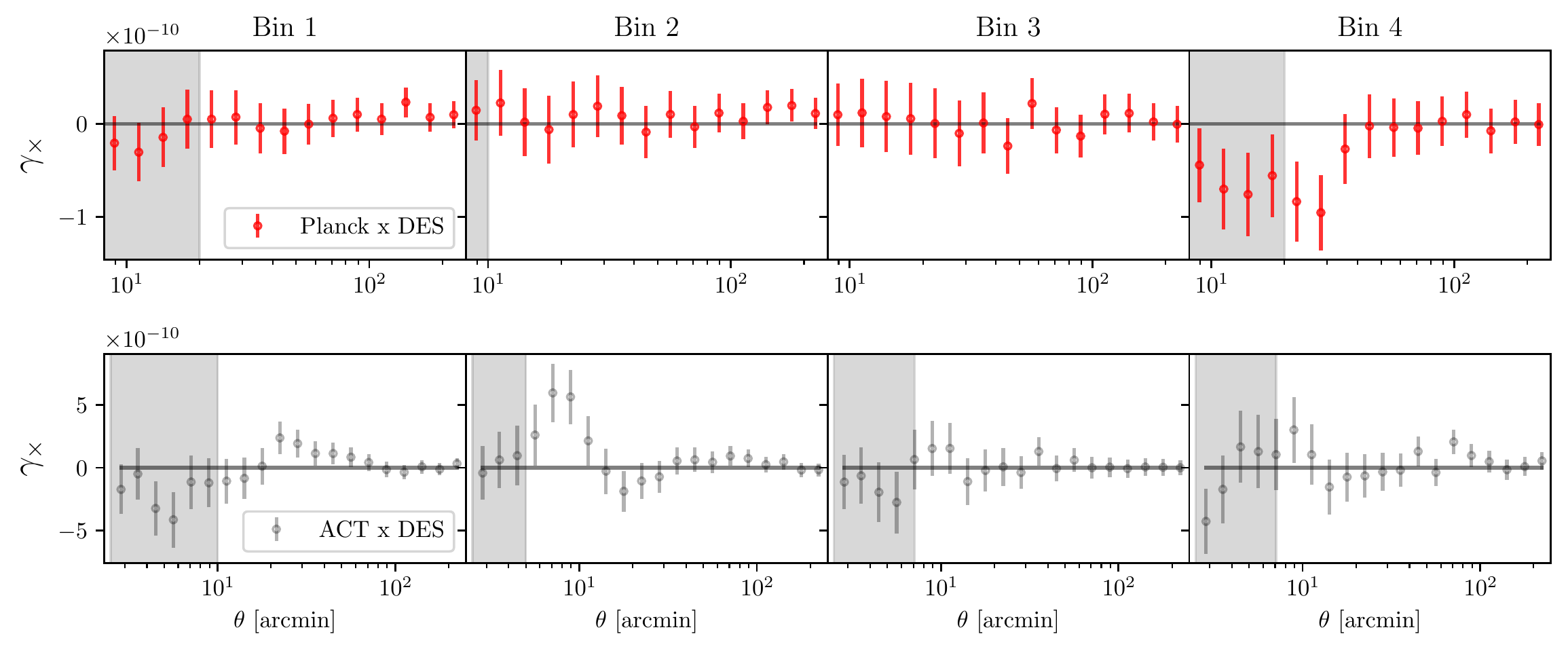}
\caption[]{Cross-component of the lensing signal around points of the Compton-$y$ maps. Grey shaded regions represent the scales removed in the main analysis. No statistically significant signal is measured (Table \ref{tests}), indicating that the null test is passed. We note that the few points at $\sim10$ arcmin for the second bin of the ACT x DES measurement that are showing a small deviation from zero are very correlated and are just a statistical fluke (we also tried to change the binning of the measurement finding again a null signal).}
\label{cross_component}
\end{figure*}

We present here a number of tests to assess whether the Compton-$y$ maps used in this work are affected by systematic effects. In particular, we are interested in the potential effect of two contaminants, namely, the cosmic infrared background (CIB) and radio point sources. 

The CIB signal is sourced by thermal emission from galaxies over a broad range of redshift ($z\sim 0.1-4$), but with the bulk of the emission mostly peaking at high redshift ($z>1$, \citet{Schmidt2015,Chiang2020}). {By assuming an effective model for the main component of the CIB that is correlated across frequency, the CIB can be de-projected from the Compton-$y$ map using the ILC method \citep{Remazeilles2011}.  We model the effective CIB SED as a modified blackbody with ``temperature'' 24 K and spectral index $\beta$ (following \cite{Madhavacheril2020}). The parameters of this effective model do not correspond to the physical SED parameters of an actual infrared source, but they do capture the frequency dependence seen in the sky-average CIB SED for the CIB halo model fit to the \textit{Planck} CIB power spectra measurements.} In this model the CIB emission rises quickly at high frequency and it is expected to mostly affect the frequency channel at 545 Ghz. We create three Compon-$y$ maps using \textit{Planck} frequency channels from 30 GHz to 545 GHz, de-projecting the CIB assuming the SED from \cite{Madhavacheril2020} with slope $\beta = 1.0,1.2,1.4$\footnote{{We also tried with a larger value of $\beta=1.6$, but found significantly increased uncertainties and no appreciable difference with respect to $\beta=1.4$.  This happens because $\beta=1.6$ is a large value which does not describe well the data, and as a consequence, the CIB de-projection does not work properly.}}. In what follows, whenever needed, we will assume the value $\beta=1.2$, and show the results for values $\beta = 1.0, 1.4$ only for comparison purposes. The value $\beta=1.2$ is the one assumed in \cite{Madhavacheril2020}, obtained by comparison with \textit{Planck} CIB halo model presented in \cite{CIBstuff}. The shear-Compton-$y$ cross-correlation $\xi^{\gamma_t y}$ signal obtained using these maps is shown in Fig.~\ref{measurement_CIB}, compared to the measurement obtained using the official \textit{Planck} NILC Compton-$y$ map presented in \cite{Planck_2016_tsz}, which does not implement any CIB de-projection. Bins 3 and 4 are the ones where the measured signal is mostly affected by the CIB de-projection procedure, depending on the value of the SED slope $\beta$. We can quantify the effect of the CIB de-projection by measuring the significance of the difference between the signal obtained using the official \textit{Planck} map and our de-projected maps. To this aim, we use a jackknife \citep{Quenouille1949,Norberg2009} covariance (see Appendix \ref{sec:covariance}) {for the measurement difference} and restrict to the scales where the jackknife estimate is not affected by the limited size of the jackknife patches ($\theta < 40$ arcmin in this case). We note that such a covariance is smaller than the measurement covariance, because sample variance should largely cancel when computing the difference between two signals measured over the same area.

This procedure is needed as the measurements involving the maps with and without CIB de-projections are highly correlated. When assuming $\beta = 1.2$, the $\chi^2$ of the difference between the signals is $\chi^2=4$, $\chi^2=12$, $\chi^2=55$, $\chi^2=110$ for 8 $d.o.f.$ for the 4 tomographic bins, respectively (Table \ref{tests}). We further generate three additional \textit{Planck} maps removing the 545 GHz frequency channel and assuming again $\beta = 1.0,1.2,1.4$. These are generally compatible with the ones including it (although they are noisier), except for the scales between $8-20$ arcminutes in bin 4 (middle panels of Fig. \ref{measurement_CIB}). Given these results, for the analysis presented in this work we decide to rely on the \textit{Planck} Compton-$y$ map obtained using all the frequency channels from 30 GHz to 545 GHz and with CIB de-projected using $\beta=1.2$ for the CIB SED. We further decide to exclude in bin 4 scales $\theta<20$ arcminutes, due to potential residual CIB contamination. Note that previous work on shear-Compton-$y$ cross-correlation using the \textit{Planck} Compton-$y$ map suggested a weaker level of CIB contamination \citep{Yan2019}. It is possible that the different redshift distribution of the galaxies used in this work could be responsible for a higher degree of CIB contamination compared to the work of \cite{Yan2019}. In particular, the redshift distributions of our bin 3 and 4 peak at higher redshift with respect to the sample used in \cite{Yan2019}, overlapping more with the high redshift galaxies responsible for the bulk of the CIB emission. Additional discussion on the effects of CIB contamination using simulated Compton-$y$ maps are provided in Appendix~\ref{sect:weird_sect}. Lastly, we also show in Fig.~\ref{measurement_CIB} the effect of CIB de-projection on the cross-correlation $\xi^{\gamma_t y}$ signal obtained using the ACT maps. The effect is negligible here, as the $\chi^2$ of the signals difference is compatible with noise: we obtain $\chi^2=4$,$\chi^2=8$, $\chi^2=14$, $\chi^2=9$ for 8 $d.o.f$,  for scales $\theta < 15$ arcmin, for the four tomographic bins (Table \ref{tests}).\footnote{These $\chi^2$ values are computed using as a covariance the measurements difference covariance, which is smaller than the covariance of a single measurement due to cancelling sample variance. As a reference, if we were to compute the $\chi^2$ of the signals difference using the single measurement covariance, we would obtain $\chi^2\sim2-3/8$ $d.o.f$} This is due to the ACT measurement being noisier than \textit{Planck} (due to the smaller area coverage) {and due to the ACT Compton-$y$ map receiving significant contributions from the ACT 98 and 150 GHz channels (even on small scales), where the CIB is relatively faint}. We also note that \cite{Schaan2021} found some evidence of mild CIB contamination for the ACT map only at scales smaller than 1 arminute, which are scales not considered in this work. Given the result of this test, in what follows, we will use the ACT map with no CIB de-projection.

As a second test, we proceed testing the potential contamination due to radio sources. Radio sources can potentially bias the signal at the 10-20\% level \citep{Shirasaki_2018}, although the exact number depends on the SED and HOD of the radio sources, which are uncertain, and on the map making algorithm. Due to the uncertainties in the SED and HOD, the radio sources contamination cannot be as easily de-projected as the CIB signal. The ACT map is created masking detected radio sources in every channel, and subsequently interpolating the map over the masked regions \citep{Madhavacheril2020}; radio sources are usually detected down to 5-10 mJy. On the other hand, the \textit{Planck} map used in this work does not have any radio sources mask applied by default. We therefore apply a radio sources mask to the \textit{Planck} Compton-$y$ map, using catalog of radio sources detected by ACT at 98 and 150 GHz. Such a catalog has not been released yet; it is built using ACT DR5 data and spans the full DES Y3 footprint \citep{Huffenberger2021}. Note that ACT can detect point-like radio sources much fainter than \textit{Planck} (1-2 orders of magnitude fainter, depending on the \textit{Planck} channel considered, \citet{Ade2014}). We masked a circular area of radius 10 arcmin around each source, and repeated the cross correlation measurement with the DES shape catalog. The masking reduced by 8\% the area available for the cross correlation. The difference in the measurements (with and without radio sources mask applied) is shown in Fig.~\ref{measurement_planck_radio}. 
No radio source signal is detected in the measurements difference: using the angular scales where the jackknife covariance is reliable, we obtained for the difference between the two signals $\chi^2/d.o.f$ = 34/32, see Table \ref{tests}. Given the lower signal-to-noise of the ACT measurement, and given the fact that this radio sources mask is already applied to the ACT map, we can also consider the impact of radio sources on the ACT map negligible.

%30.217024901182207 28 %Note that the signal with the radio sources masked is has a preferentially lower amplitude compared 

%Note that using such an aggressive masking we are also preferentially removing regions of the map where a positive is signal is expected, rather than removing the true radio source contamination.

As a last systematic test, we checked the cross-component of the mean shear around every point of the Compton-$y$ maps. The cross-component is a standard null test in galaxy-galaxy lensing studies, as it should be compatible with zero if the shear is produced by gravitational lensing alone. The cross-component should also vanish in the presence of systematic effects that are invariant under parity. We test this in Fig.~\ref{cross_component}. Using a theory covariance, we obtain $\chi^2/d.o.f$ = 84/68 and $\chi^2/d.o.f$ = 93/80 for the \textit{Planck} x DES and ACT x DES measurements (Table \ref{tests}), respectively, indicating compatibility with a null signal and that the null test is passed.

%0.6532663316582915
%0.09095208777240517 1.3349149763773749 84.02219267026624 68
%/global/homes/m/mgatti/.conda/envs/py3s/lib/python3.6/site-p%ackages/ipykernel_launcher.py:95: DeprecationWarning: %elementwise comparison failed; this will raise an error in %the future.
%0.592964824120603
%0.13634824922913652 1.0968739483412056 93.95422424363593 80

%Before using our measurements to test different feedback models, we performed a number of null tests. In particular, we tested that our measurements are robust against a number of potential contaminants: the cosmic infrared background (CIB), and radio sources. For the measurement involving the \textit{Planck} data, we also compared the measurement obtained using two different versions of the Compton-$y$ map (based on the \texttt{MILCA} and {NILC} algorithms). For both \textit{Planck} x DES and ACT x DES measurements, we also measured the cross-component of the mean shear around every point of the Compton-$y$ maps, which should vanish if the signal is produced by gravitational lensing alone (or in presence of systematic effects that are invariant under parity). None of these tests flagged anything problematic in our data. The results of these tests are shown and  discussed in Appendix \ref{sect:map_tests}.

\subsection{{Tests of feedback models}}\label{sect:feedback_models}

\begin{figure*}
\includegraphics[width=1.\textwidth]{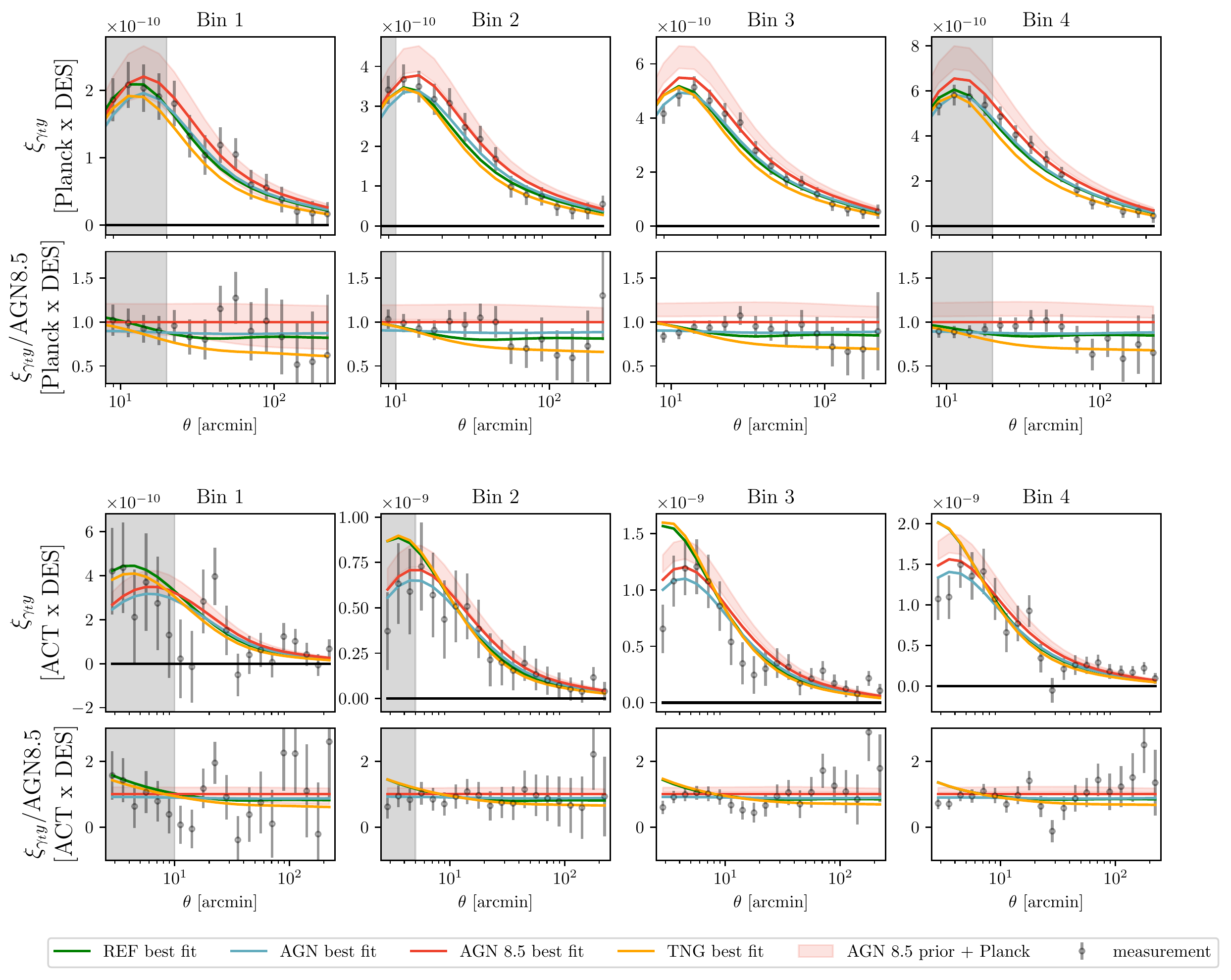}
\caption[]{Measured shear-Compton-$y$ cross-correlation $\xi^{\gamma_t y}$ and best fit models for four feedback models (TNG, AGN, AGN8.5 and REF). These models have been obtained assuming the re-scaled NFW profile for the lensing signal. Top panels show the cross correlations between \textit{Planck} and DES Y3 data for the four tomographic bins, whereas bottom panels show the correlations between ACT and DES Y3 data. As a comparison, we also show as a coloured band encompassing the 68\% confidence interval of the prior for the AGN 8.5 scenario. The grey shaded regions represent the angular scales not included in our analysis (see Table~\ref{table_scale_cuts}.)}
\label{final_result}
\end{figure*}

\begin{table}
%\tiny
\caption {The MAP value of $\chi^2$ for the feedback models, obtained assuming \textit{Planck} priors on $\sigma_8$ and $\Omega_{\rm m}$, and marginalising over nuisance parameters as explained in \S~\ref{sect:like}. The top half of the table refers to the models obtained re-scaling the NFW profile for the lensing signal; on the other hand, the bottom half of the table refers to the analysis where we model the lensing signal using the Mead model. We also report the update-difference-in-mean (UDM) tension for the best fit models with respect to their priors.}
\centering
%\begin{adjustbox}{width=0.7\textwidth}

\begin{tabular}{|c|c|c|c|c|c|}
 \hline
  \multicolumn{6}{|c|}{\,\, \, \,\,    \,\,       \textit{Planck} prior (NFW re-scaling)}    \\
 \hline
 & \textbf{B12} & \textbf{AGN} & \textbf{AGN 8.5} & \textbf{REF} &  \textbf{TNG} \\

% \hline
%bin 1 & & 49 & 49 & 48 & & 44 & 47 & 46 & &22 \\
%bin 2 & & 61 & 67 & 61 & & 53 & 59 & 57 & &29 \\
%bin 3 & & 79 & 101 & 84 & & 72 & 72 & 83 & &35 \\
%bin 4 & & 70 & 97 & 77 & & 61 & 62 & 73 & &35 \\

%(191/122)
%(7.4 $\sigma$)

%planck_B12_AGN 154
%planck_LB_REF 155
%planck_LB_AGN 155
%planck_LB_AGN85 154
%planck_TNG 154

 \hline

$\chi^2/d.o.f.$ &   \,\,\,\,\,\,\,\,-\,\,\,\,\,\,\,\, & 172/119 & 158/119 & 189/119 & 198/119 \\
UDM tension &\,\,\,\, \,\,\,\,-\,\,\,\,\,\,\,\, & 4.5 $\sigma$ & 2.2 $\sigma$ & 3.3 $\sigma$ &  4.3 $\sigma$ \\
\hline
\end{tabular}

\begin{tabular}{|c|c|c|c|c|c|c|}
 \hline
  \multicolumn{6}{|c|}{\,\, \, \,\,    \,\,       \textit{Planck} prior (free $A_{\rm Mead}$,$\eta_{\rm Mead}$) }    \\
 \hline
 & \textbf{B12} & \textbf{AGN} & \textbf{AGN 8.5} & \textbf{REF}& \textbf{TNG} \\

% \hline
%bin 1 & & 49 & 49 & 48 & & 44 & 47 & 46 & &22 \\
%bin 2 & & 61 & 67 & 61 & & 53 & 59 & 57 & &29 \\
%bin 3 & & 79 & 101 & 84 & & 72 & 72 & 83 & &35 \\
%bin 4 & & 70 & 97 & 77 & & 61 & 62 & 73 & &35 \\
 \hline

$\chi^2/d.o.f.$ &   154/118 & 155/118 & 154/118 & 155/118 &154/118  \\
UDM tension &  0.7 $\sigma$ & 1.5 $\sigma$ & 1.4 $\sigma$ & 0.5 $\sigma$  & 0.3 $\sigma$  \\
\hline
\end{tabular}

%\end{adjustbox}
\label{significance_f}
\end{table}

\begin{figure}
\includegraphics[width=0.45\textwidth]{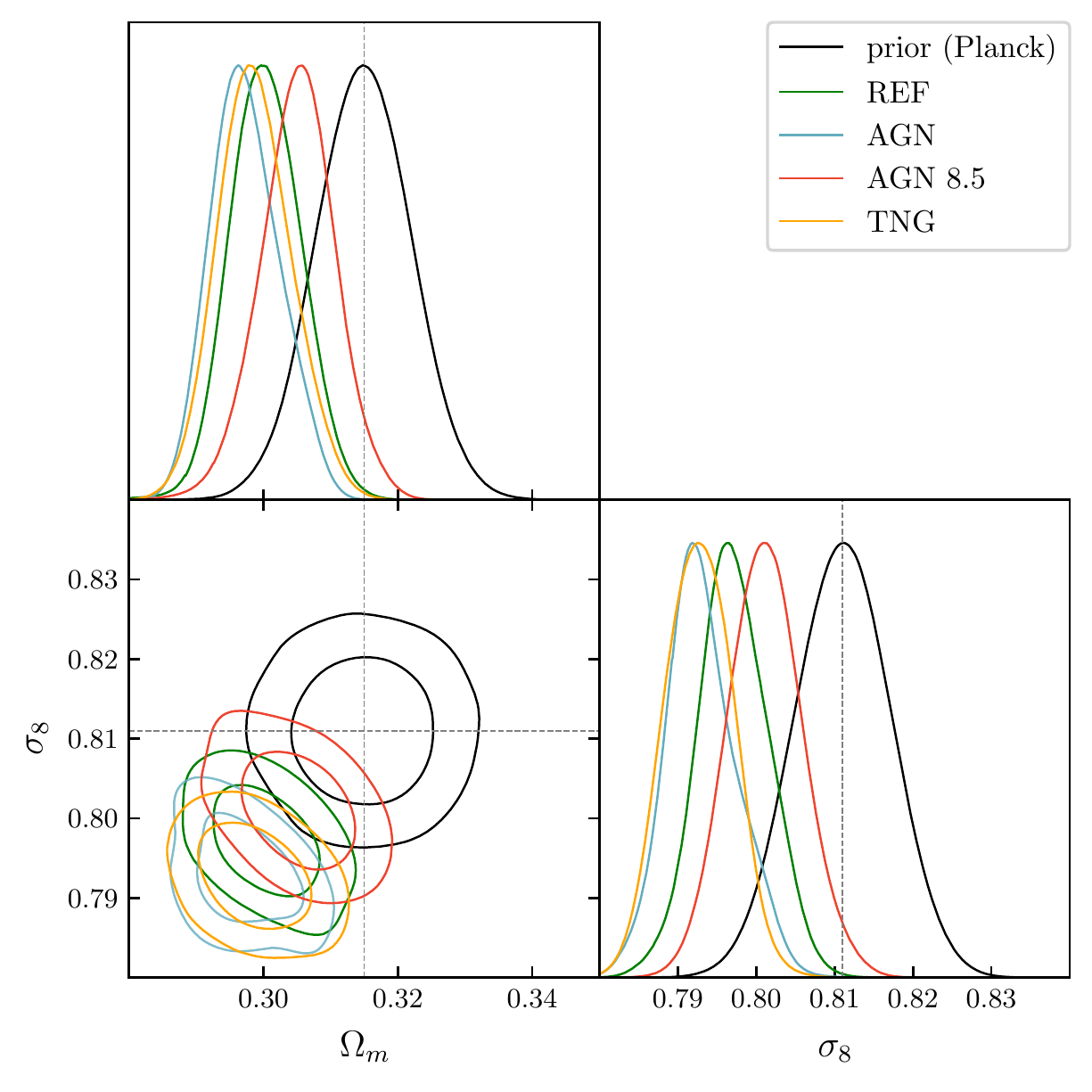}
\includegraphics[width=0.45\textwidth]{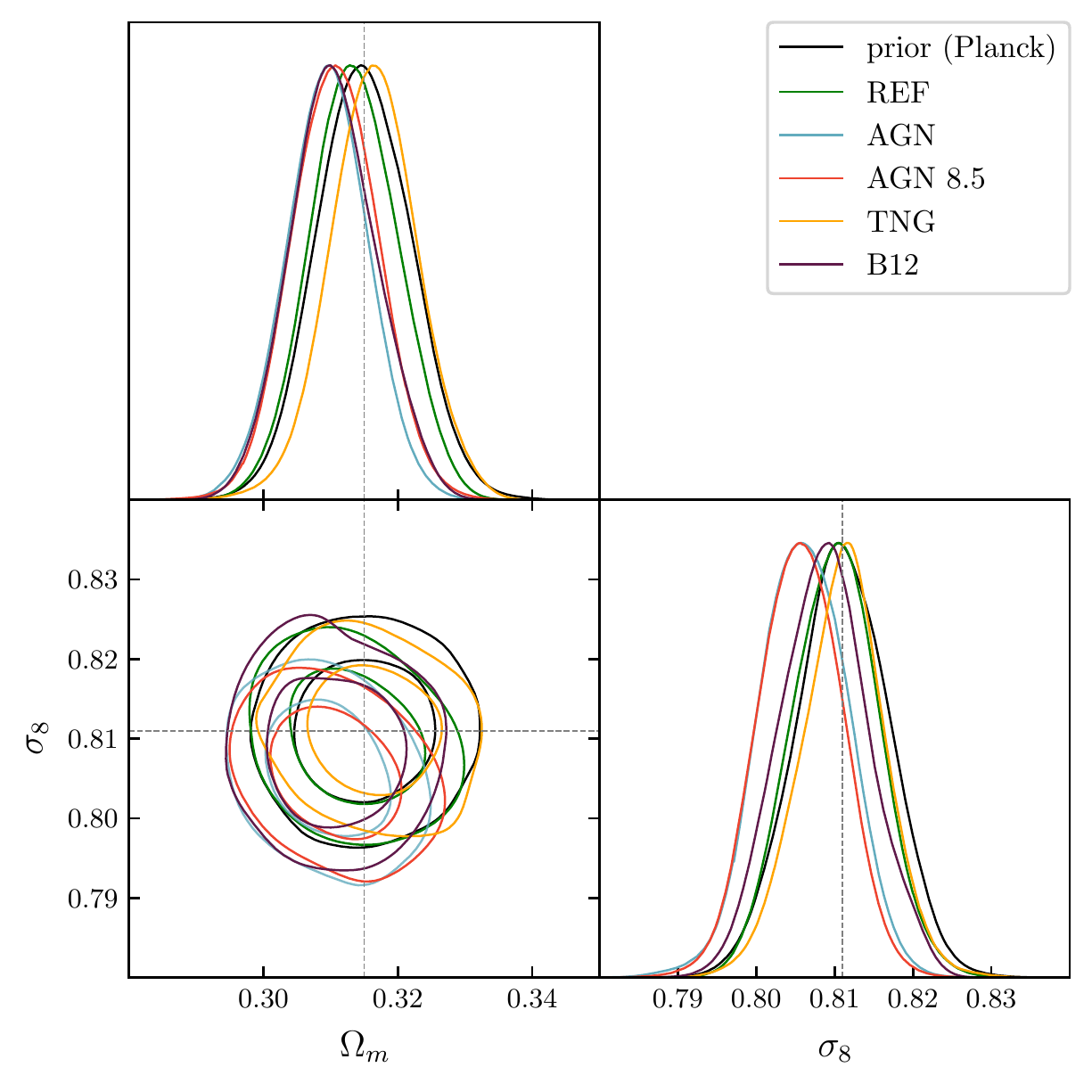}

\caption[]{Posterior for the parameters $\sigma_8$ and $\Omega_m$ for the four (five) feedback models: (B12) , REF, AGN, AGN 8.5 and TNG. We also show the \textit{Planck} prior centered in each plot. Top refers to the case where we used the NFW re-scaling to model the lensing part of our signal; the bottom plot refers to the more conservative analysis where we used the Mead model. The marginalised contours in these figures show the 68\% and
95\% confidence levels. When the NFW re-scaling is used, the data are in mild to moderate tension with the \textit{Planck} prior (2.2$\sigma$-4.5$\sigma$, see Table \ref{significance_f}).}
\label{contour_plots}
\end{figure}

\begin{figure}
\includegraphics[width=0.45\textwidth]{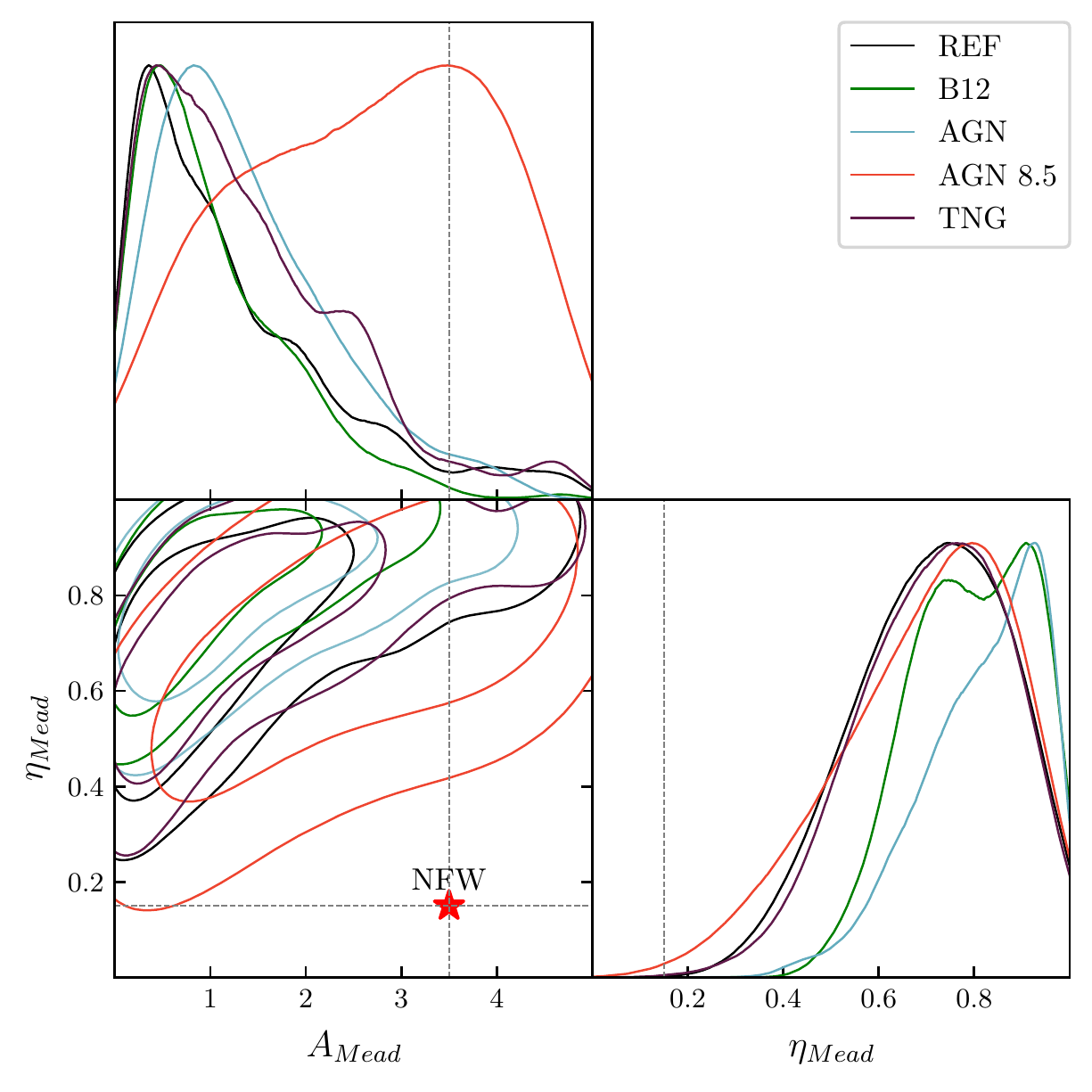}

\caption[]{Posterior for the Mead model parameters $A_{\rm Mead}$ and $\eta_{\rm Mead}$ for five feedback models: B12 , REF, AGN, AGN 8.5 and TNG. The ``star'' indicates the values of the Mead models that best describes a NFW profile. The marginalised contours in this figure shows the 68\% and
95\% confidence levels.}
\label{contour_plots_mead}
\end{figure}

\begin{figure*}
\includegraphics[width=0.9\textwidth]{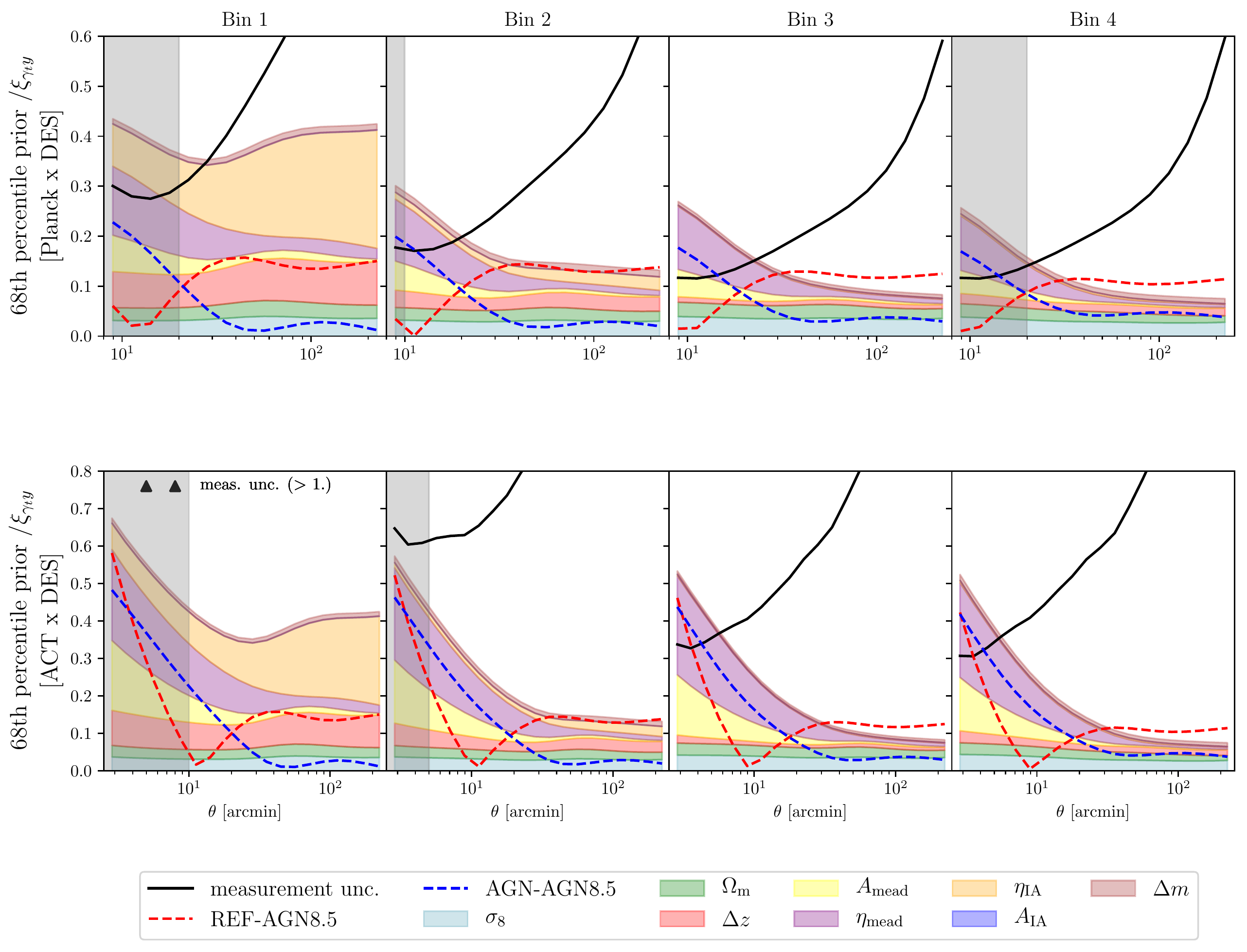}
\caption[]{The plot assesses how our ability to discriminate among different feedback models is limited by our measurement uncertainties and nuisance parameters. In particular, the coloured bands show the contribution of each single nuisance parameter to the total 68\% confidence interval of our prior. We assume \textit{Planck} priors for $\sigma_8$ and $\Omega_{\rm m}$. The black line represents the measurement uncertainty. The vertical grey line indicates the scale cut adopted in this analysis (Table~\ref{table_scale_cuts}). Quantities in the figure are divided by a theory data vector $\xi^{\gamma_t y}$ obtained for the AGN8.5 feedback model. We also show the fractional difference with respect to two other models: REF and AGN. Areas where the black line is below the combined prior budget can be used to differentiate between models.}
\label{priors_models}
\end{figure*}

After characterising the measurement, we compare it to theoretical predictions using the halo model framework and pressure profiles as estimated from a number of hydrodynamical simulations. The pressure profiles and feedback models considered in this section are the five models introduced in \S~\ref{sect:compton-y_theo} (B12, REF, AGN, AGN8.5, TNG). For the lensing part of the signal, we both model our theoretical predictions using the re-scaled NFW profiles and using the Mead model (\S~\ref{sect:shear_signsl}). We compare our theoretical models to the measurement marginalising over a number of nuisance parameters modelling astrophysical and measurement systematics, including photometric redshift uncertainties, intrinsic alignment and shear calibration biases, as described in \S~\ref{sect:like}. We also marginalise over $\sigma_8$ and $\Omega_{\rm m}$ assuming \textit{Planck} priors for the cosmological parameters (but we also repeat the analysis in Appendix~\ref{sec:DES_prior} using DES Y1 priors). When using the Mead model for the lensing kernel, we also marginalise over Mead model parameters. Such marginalisation over systematics has generally been neglected in early works on weak lensing - tSZ correlations (but see \citet{Osato2020}). When comparing models to the measurements, we always jointly fit the ACT x DES and \textit{Planck} x DES measurements. 

We start considering the analysis with re-scaled NFW profiles for the lensing part of the signal. This is the most constraining setup, as we do not marginalise over the effect of baryonic feedback on the matter profile, but rather we rely on the measurements of power spectra from hydrodynamic simulations to re-scale our theoretical predictions of the matter profile. Note that such re-scaling has a milder impact on the measurement compared to the effect of differences in the pressure profile from different hydrodynamical simulations. The main caveat of this approach is that such re-scaling neglects any halo mass dependency of the effect of baryonic feedback on the matter profile, as the power spectra do not carry an information about the halo mass, so it can be seen as an ``effective'' re-scaling. Note that due to the lack of the 3D power spectrum needed for the re-scaling for the B12 model, we did not consider it in this first part of the analysis.

Fig.~\ref{final_result} shows the \textit{Maximum A Posteriori},  (MAP) models obtained by assuming the four feedback models and marginalising over nuisance and cosmological parameters. {The pressure profiles are not varied, only the nuisance and cosmological parameters; so the best-fit models for our measurements are obtained using the pressure profiles from the hydrodynamical simulations and the best-fit nuisance and cosmological parameters from the analysis.} The data likelihood at MAP are approximately described as a $\chi^2$ distribution with $d.o.f.$ equal to the number of data points $N_{\rm points}$ minus the effective number of parameters $N_{\rm eff}$ constrained by our data compared to the priors it began with \citep{Raveri2019}. In our case $N_{\rm points} =123$, whereas the effective number is:
\begin{equation} \label{Eq:NumberEffectiveParameters}
N_{\rm eff} = N -{\rm tr}[(\mathcal{C}^{\rm p })^{-1}\mathcal{C}^{\rm p +d}],
\end{equation}
where $\mathcal{C}^{\rm p }$ is the covariance of the prior, and $\mathcal{C}^{\rm p + d }$ is the covariance of the prior updated by the data (i.e., the posterior). Moreover, $N=12$, that is, the number of free 12 parameters in our analysis. We obtain, depending on the model, $N_{\rm eff} \sim 4$. We note that although we free 12 parameters in our analysis, some of them are tightly constrained by their prior, such that $N_{\rm eff}< 12$. The data likelihood at MAP is reported in Table \ref{significance_f}. The model that provides the MAP best fit $\chi^2$ is the AGN 8.5 model, followed by the AGN 8, REF and the TNG model. In particular, the last two feedback models are penalised by the comparison with the ACT x DES data, which prefer the scenarios with a lower amplitude of the pressure profile at small scales, compatible with the ejection of gas from the inner part of the halo. On the other hand, the AGN 8.5 model also provides a better fit to the \textit{Planck} x DES measurement at all scales (especially for bins 3 and 4) compared to the AGN model, hence providing the best $\chi^2$ among all the scenarios probed here.

We compare if the best fit models are in tension with their priors. In order to quantify the level of agreement/disagreement we use a Gaussian estimator, called update difference-in-mean (UDM) statistic \citep{Raveri2019}. The UDM statistics compare the mean parameters from the prior $\hat{\theta}^{\rm p}$ with the updated values $\hat{\theta}^{\rm p + d}$ obtained running the analysis on data. {This statistic assumes either flat or Gaussian priors (which is satisfied for all the parameters considered in this analysis, see Table~\ref{table1}), and requires the posterior of the well-constrained parameters to be approximately Gaussian (which is also satisfied)}. In particular, we can define

%This is evident in Fig.~\ref{final_result} for the \textit{Planck} prior case: for the two highest redshift bins, all the best fit models are well below their priors. 
%
\begin{equation}
Q_{\rm UDM} = (\hat{\theta}^{\rm p + d }-\hat{\theta}^{\rm p})^T \left( \mathcal{C}^{\rm p} -\mathcal{C}^{\rm p + d } \right)^{-1}(\hat{\theta}^{\rm p + d }-\hat{\theta}^{\rm p}),
\end{equation}
where the difference in the mean of the parameters $(\hat{\theta}^{\rm p + d }-\hat{\theta}^{\rm p})$ is weighted by the parameters inverse covariance. If the parameters are Gaussian distributed, $Q_{\rm UDM}$ is chi-squared distributed with ${\rm rank}(\mathcal{C}^{\rm p + d } -\mathcal{C}^{\rm p})$ degrees of freedom. The UDM tension is reported in Table~\ref{significance_f}. A tension would imply that for a \textit{Planck} cosmological model, the tSZ signal that we measure is in tension with the predictions of these feedback models. For the AGN, REF and TNG models, the best fit are in tension at the 3-4 $\sigma$ level with their priors, with the best-fit models to the data preferring different values of the cosmological parameters $\sigma_8$ and $\Omega_{\rm m}$ than the ones measured by \textit{Planck}. This is also shown in Fig.~\ref{contour_plots}, which reports the posterior of the $\sigma_8$ and $\Omega_{\rm m}$ parameters, compared to the priors used in the analysis.\footnote{We caution the reader from ``reading'' the exact value of the UDM tension from Fig.~\ref{contour_plots}: the posteriors are the results of the combination of the \textit{Planck} prior and the feedback models likelihood, whereas the UDM tension computes the tension between the model \textit{alone} and the prior, which is in general larger than what it could be inferred ``by-eye'' from Fig.~\ref{contour_plots}.}
On the other hand, the AGN 8.5 scenario, which is also the one the provides the best fit to the data, is not in significant tension with its prior. Given the fact that the AGN 8.5 scenario is also the one that provides the best-fit $\chi^2$, this reinforces the idea that the data prefers a model with a low amplitude of the pressure profile at small scales.

%We remind the reader that in this work we do not attempt to vary the parameters of the pressure profiles, but we assume the values calibrated against hydrodynamical simulations. 

%In particular, it can be noted that all the models are in tension ($> 4 \sigma$) with the \textit{Planck} prior. These values quantify the disagreement between the best fit models and the prior bands for bin 3 and 4 shown in Fig.~\ref{final_result}. Interpreting at face values these numbers, they indicate that either the data suggests different values of the cosmological parameters $\sigma_8$ and $\Omega_{\rm m}$ than the ones measured by \textit{Planck}, or that all the feedback models fail capturing a redshift evolution of the pressure profiles, at the $>4\sigma$ level. This tension can also be noted in Fig.~\ref{contour_plots}, where we show the posterior of the $\sigma_8$ and $\Omega_{\rm m}$ parameters, compared to the priors used in the analysis. We remind the reader that in this work we do not attempt to vary the parameters of the pressure profiles, but we assume the values calibrated against hydrodynamical simulations. In \paperB, however, we demonstrate that when the parameters of the pressure profiles are varied, assuming a \textit{Planck} cosmology, the data indeed prefer a redshift evolution of the pressure profile amplitude in the redshift range $z \sim 0.3-0.5$ different than the ones predicted by the feedback models considered here.

We then repeat the analysis using the Mead model for the lensing signal, instead of using the re-scaled NFW profiles. This approach is more conservative, as we let the data re-scale the lensing profile instead of relying on the power spectra measured on simulations. This approach is, however, less constraining, as we marginalise over the Mead model parameters using wide priors. In this case all the models provide a similar best fit $\chi^2$. The fact that the best fit of all the different feedback models are similar is related to the large, uninformative prior on the Mead model parameters, which, together with the freedom allowed by the priors on the cosmological and nuisance parameters, absorbs most of the differences between models. Interestingly, the best fit $\chi^2$ of all the models obtained freeing Mead model parameters is not too different from the  best fit $\chi^2$ obtained for the AGN 8.5 model and re-scaling the NFW profile.

{We show in Fig.~\ref{contour_plots_mead} the posterior of the Mead model parameters for each feedback models. In general, the feedback models prefer smaller $A_{\rm Mead}$ and larger $\eta_{\rm Mead}$ than the NFW profile (except for the $A_{\rm Mead}$ parameter for the AGN 8.5 feedback model), which implies less concentrated and more ``bloated'' halos}. Lastly, we note that the UDM statistics for this second analysis is not in tension with the \textit{Planck} prior (Fig.~\ref{contour_plots} and Table~\ref{significance_f}), owing to a larger prior space.

{We test whether our findings are robust against the exact value of the parameter $\beta$ used to de-project the CIB contribution in the \textit{Planck} map. To test this, we ran a full analysis using the measurements without CIB de-projection for the \textit{Planck} x DES measurement. This is an extreme case, as we believe the CIB de-projection is necessary, and the value $\beta=1.2$ used in the fiducial analysis is justified by \cite{Madhavacheril2020}. Nonetheless, we confirm that also in this unrealistic case, when running using the re-scaled NFW profiles the data prefers the AGN 8.5 model in terms of best fit $\chi^2$. We find, however, that the UDM tension metric increases for all the models\footnote{In particular, when assuming NFW re-scaling we obtain an UDM tension of $6.3\sigma$, $3.8\sigma$, $5.2\sigma$, $6.1\sigma$ for the AGN, AGN 8.5, REF and TNG models, respectively. The UDM tension increases also when the Mead model is assumed: $2.3 \sigma$, $2.6\sigma$, $1.2\sigma$, $1.4\sigma$, $2.0 \sigma$ for the AGN, AGN 8.5, REF, TNG and B12 models, respectively.}: neglecting the CIB de-projection makes our analysis prefer a different cosmology than \textit{Planck}, but it does not have an impact on the feedback model selection. }

Finally, we investigate the limitation in our ability to constrain different feedback models due to our measurement uncertainties and lack of tight priors on our nuisance parameters. This is shown in Fig.~\ref{priors_models}. The coloured bands show the contribution of each single nuisance parameter to the total 68\% confidence interval of our prior. Note that we assumed the \textit{Planck} prior for $\sigma_8$ and $\Omega_{\rm m}$, and we considered the conservative scenario where we also marginalise over Mead model parameters.  Quantities in Fig.~\ref{priors_models} are shown with respect to a theory data vector obtained assuming the AGN8.5 model. We also show in the plot the fractional difference (in absolute value) with respect to two other feedback models (AGN, REF). 

For the purposes of feedback model selection, the ideal situation would be a regime where differences between models are larger than measurement uncertainties (black lines), and that measurement uncertainties are not sub dominant with respect to the prior. This is not happening with our current data: our measurement uncertainties are generally larger than the difference between the models, except for a small window at $\sim 10 - 20$ arcmin in bin 3 and 4 for the \textit{Planck} x DES measurement and small scales ($<5$ arcmin) in bin 3 and 4 for the ACT x DES measurement. In this respect, future releases of the ACT Compton-$y$ map \citep{Naess2020} will definitely improve the situation. The next ACT Compton-$y$ map will cover the full DES Y3 footprint. This means that our measurement uncertainties from the ACT x DES correlations will be significantly smaller compared to the ones shown in this paper - as a comparison, they should be smaller than the ones quoted for \textit{Planck} x DES in Fig.~\ref{priors_models} (as in this work we removed a part of the \textit{Planck} data), with the measurement extending down to small scales ($\sim2.5$ arcmin).

In Fig.~\ref{priors_models} we note the large contribution of the Mead model nuisance parameters to the 68\% confidence interval of the prior. This is a substantial contribution that dominates the prior at scales $<20$ arcminutes, and it explains why when marginalising over Mead model parameters our data could not discriminate among feedback models. At low redshift, uncertainties in the redshift estimates and intrinsic alignment parameter $\eta_{\rm IA}$ are also providing a substantial contribution {(note that $A_{\rm IA}$ is instead tightly constrained by the prior, so it has a negligible impact and its 68\% confidence interval cannot be seen in Fig.~\ref{priors_models})}. {This is expected as the lensing signal is smaller in amplitude at low redshift compared to higher redshift, and uncertainties on intrinsic alignment or redshift estimates can have a larger impact}. When marginalising over the Mead model parameters, we did not assume any tight prior. In principle, one could estimate the correct values of $A_{\rm Mead}$ and $\eta_{\rm Mead}$ measuring the matter profiles in hydrodynamical simulations for the range of halo masses our measurement is sensitive to (which should be more accurate than just re-scaling the NFW profiles). Additional constraints on the Mead model parameters can be provided by a joint analysis with cosmic shear, including small scales (although cosmic shear is sensitive to lower mass halos compared to $\xi^{\gamma_t y}$). When varying Mead model parameters, we are not assuming any prior on the relation with the pressure profiles parameters; in principle, however, the two might be related. As we mentioned in \S~\ref{sect:shear_signsl}, since our Mead model is an effective model, it is hard to place physically motivated priors on the relation between the pressure profile parameters and the Mead model parameters. A joint cosmic shear and tSZ-shear analysis would benefit from having physically motivated priors relating the two sets of parameters, as this would help tighten the constraints. In this respect, more coherent frameworks (as the one introduced in \citet{Mead2020b}) where the shear and tSZ signals are modelled starting from the distribution of gas, matter, and stars, might be better suited to this task.  Future analysis should also improve the modelling of intrinsic alignment; in this work we took a conservative approach and we removed a good portion of angular scales from the two lowest redshift bins due to the uncertainties in the modelling of the 1-halo IA contribution due to satellite galaxies alignment. A joint analysis with cosmic shear, with better IA modelling on small scales, could allow us to also use the smallest scales of our measurements at low redshift, with tighter constraints on IA nuisance parameters.

%\subsection{Tests on 1-halo 2-halo transition regime}
%
%discuss why the best fit models can't fit well the transition regime.
%show the impact of alpha. show chi2 excluding those scales. discuss %non linear halo bias. what else?

\section{Summary} \label{sec:summary}
This is the first of two works on cross-correlations between thermal Sunyaev-Zeldovich (tSZ) maps from \textit{Planck} and the Atacama Cosmology Telescope (ACT) and weak gravitational lensing shears measured during the first three years of observations of the Dark Energy Survery (DES). {This correlation is sensitive to the thermal energy in baryons as a function of redshift, and is in principle a powerful probe of astrophysical feedback.} In this work we presented the cross-correlation measurements: we detected the {correlation} at a significance of 21$\sigma$, the highest significance {to date}. We also presented a series of systematic tests, where we tested the effect of potential contaminants on our measurements, including cosmic infrared background (CIB) and radio sources. {We found that CIB has a substantial effect on the \textit{Planck} x DES measurement, whereas the ACT x DES measurement was not significantly affected, {probably due to the ACT Compton-$y$ map receiving significant contributions from the ACT 98 and 150 GHz channels, where the CIB is relatively faint}, and also due to the noisier nature of the latter measurement. In order to account for the CIB effect, we built a CIB de-projected Compton-$y$ map for the \textit{Planck} data and used it in our main analysis}. 

We then used the shear-$y$ correlation measurements to test a number of different feedback models, modelling the correlations using the halo model formalism. In particular, we modelled the tSZ part of the signal using a number of different pressure profiles calibrated against hydrodynamical simulation which have implemented different baryonic feedback models. On the other hand, the shear part was modelled either using a re-scaled NFW profile or implementing the Mead halo model \citep{Mead2015}. In the first approach, the NFW profile was re-scaled by a mass-independent factor given by the ratio of the power spectrum from a dark-matter only simulation and the power spectrum from a hydrodynamical simulation with dark-matter and a sub-grid prescription for baryonic effects. In the second approach, the lensing kernel was modelled by a generalised NFW profile \citep{Mead2015} with extra degrees of freedom to take into account the effect of baryonic feedback processes. When comparing our models to our measurement, we kept the pressure profile model fixed, as our goal was to discriminate among different feedback models. Note that a different approach where the pressure profile parameters are varied is adopted in \paperB. 

In our analysis, we marginalised over 10 nuisance parameters capturing redshift uncertainties, shear calibration biases, and intrinsic alignment effects adopting DES priors. We also marginalised over $\Omega_{\rm m}$ and $\sigma_8$ using \textit{Planck} and DES priors, and when used, over the Mead halo model parameters. 
We found when using the re-scaled NFW profile in combination with the pressure profiles from hydrodynamical simulations, the data preferred a lower amplitude of the pressure profile at small scales, compatible with a scenario with stronger AGN feedback and ejection of gas from the inner part of the halos (the AGN 8.5 model). We quantified the level of agreement/disagreement of each model with the data using Gaussian estimators \citep{Raveri2019}, and we found that, when assuming \textit{Planck} priors on the cosmological parameters $\Omega_{\rm m}$ and $\sigma_8$, all the models were in $3-4 \sigma$ tension with the prior, except for the AGN 8.5 model, which showed a lower tension ($2.2\sigma$). This means that for a \textit{Planck} cosmological model, the tSZ signal that we measure is in tension with the predictions of most of these feedback models (except for the AGN 8.5 model). When using the Mead model in combination with the pressure profiles from hydrodynamical simulations, we obtained weaker constraints due to the extra nuisance parameters of the model, for which we did not assume any tight prior. In this case, the data could not discriminate between different baryonic prescriptions, but generally preferred halos less concentrated and more bloated compared to a NFW profile. 

We then discussed whether the lack of tight priors on the nuisance parameters is limiting our analysis, finding that the Mead model parameters are dominating our prior volume. We discussed how one could place tighter constraints on the Mead model parameters measuring the matter profiles in hydrodynamical simulations for the range of halo masses to which our measurement is sensitive.  Additional constraints on the Mead model parameters could also be provided by a joint analysis with cosmic shear, including small scales - possibly with a more coherent, physically motivated framework, as the one introduced in \citet{Mead2020b}. In general, it might also be useful to include in future analyses additional correlations sensitive to different halo masses (e.g., Compton-$y$ - galaxy cross-correlations or Compton-$y$ auto-correlations), in order to be able to study feedback mechanisms over a wide range of halo masses. Last, we mentioned how future data and in particular future releases of the ACT Compton-$y$ map (which will cover the full DES footprint) will improve our ability to discriminate between different feedback models, as these maps will allow us to measure with high accuracy the Compton-$y$ shear correlation at small scales, where feedback models show a large variance. This will constitute a substantial improvement over current ACT data (which have a limited overlap with DES data and hence noisier cross-correlation measurements) and over \textit{Planck} data, which have a low angular resolution that does not allow us to efficiently probe the small scale regime.

\section*{Data Availability}
The full \mcal\ catalogue will be made publicly available following publication, at the URL \url{https://des.ncsa.illinois.edu/releases}. The code used to perform the tests in this manuscript will be made available upon reasonable request to the authors.

\section*{Acknowledgements}
This paper has gone through internal review by the DES and ACT collaborations. SP is supported in part by the US Department of Energy Grant No. DE-SC0007901 and NASA ATP Grant No. NNH17ZDA001N. ES is supported by DOE grant DE-AC02-98CH10886. KM acknowledges support from the National Research Foundation of South Africa. ZX is supported by the Gordon and Betty Moore Foundation. JPH acknowledges funding for SZ cluster studies from NSF AAG number AST-1615657.

Funding for the DES Projects has been provided by the U.S. Department of Energy, the U.S. National Science Foundation, the Ministry of Science and Education of Spain, 
the Science and Technology Facilities Council of the United Kingdom, the Higher Education Funding Council for England, the National Center for Supercomputing 
Applications at the University of Illinois at Urbana-Champaign, the Kavli Institute of Cosmological Physics at the University of Chicago, 
the Center for Cosmology and Astro-Particle Physics at the Ohio State University,
the Mitchell Institute for Fundamental Physics and Astronomy at Texas A\&M University, Financiadora de Estudos e Projetos, 
Funda{\c c}{\~a}o Carlos Chagas Filho de Amparo {\`a} Pesquisa do Estado do Rio de Janeiro, Conselho Nacional de Desenvolvimento Cient{\'i}fico e Tecnol{\'o}gico and 
the Minist{\'e}rio da Ci{\^e}ncia, Tecnologia e Inova{\c c}{\~a}o, the Deutsche Forschungsgemeinschaft and the Collaborating Institutions in the Dark Energy Survey. 

The Collaborating Institutions are Argonne National Laboratory, the University of California at Santa Cruz, the University of Cambridge, Centro de Investigaciones Energ{\'e}ticas, 
Medioambientales y Tecnol{\'o}gicas-Madrid, the University of Chicago, University College London, the DES-Brazil Consortium, the University of Edinburgh, 
the Eidgen{\"o}ssische Technische Hochschule (ETH) Z{\"u}rich, 
Fermi National Accelerator Laboratory, the University of Illinois at Urbana-Champaign, the Institut de Ci{\`e}ncies de l'Espai (IEEC/CSIC), 
the Institut de F{\'i}sica d'Altes Energies, Lawrence Berkeley National Laboratory, the Ludwig-Maximilians Universit{\"a}t M{\"u}nchen and the associated Excellence Cluster Universe, 
the University of Michigan, the National Optical Astronomy Observatory, the University of Nottingham, The Ohio State University, the University of Pennsylvania, the University of Portsmouth,  SLAC National Accelerator Laboratory, Stanford University, the University of Sussex, Texas A\&M University, and the OzDES Membership Consortium.

Based in part on observations at Cerro Tololo Inter-American Observatory at NSF's NOIRLab (NOIRLab Prop. ID 2012B-0001; PI: J. Frieman), which is managed by the Association of Universities for Research in Astronomy (AURA) under a cooperative agreement with the National Science Foundation.

The DES data management system is supported by the National Science Foundation under Grant Numbers AST-1138766 and AST-1536171. The DES participants from Spanish institutions are partially supported by MINECO under grants AYA2015-71825, ESP2015-66861, FPA2015-68048, SEV-2016-0588, SEV-2016-0597, and MDM-2015-0509, 
some of which include ERDF funds from the European Union. IFAE is partially funded by the CERCA program of the Generalitat de Catalunya. Research leading to these results has received funding from the European Research Council under the European Union's Seventh Framework Program (FP7/2007-2013) including ERC grant agreements 240672, 291329, and 306478. We  acknowledge support from the Brazilian Instituto Nacional de Ci\^encia e Tecnologia (INCT) e-Universe (CNPq grant 465376/2014-2).

This manuscript has been authored by Fermi Research Alliance, LLC under Contract No. DE-AC02-07CH11359 with the U.S. Department of Energy, Office of Science, Office of High Energy Physics.

Support for ACT was through the U.S. National Science Foundation through awards AST-0408698, AST-0965625, and AST-1440226 for the ACT project, as well as awards PHY-0355328, PHY-0855887 and PHY-1214379. Funding was also provided by Princeton University, the University of Pennsylvania, and a Canada Foundation for Innovation (CFI) award to UBC. ACT operates in the Parque Astronomico Atacama in northern Chile under the auspices of the Agencia Nacional de Investigacion y Desarrollo (ANID).The development of multichroic detectors and lenses was supported by NASA grants NNX13AE56G and NNX14AB58G. Detector research at NIST was supported by the NIST Innovations in Measurement Science program.

%%%%%%%%%%%%%%%%%%%%%%%%%%%%%%%%%%%%%%%%%%%%%%%%%%



%\bibliographystyle{mnras}


%%%%%%%%%%%%%%%%%%%%%%%%%%%%%%%%%%%%%%%%%%%%%%%%%%

%%%%%%%%%%%%%%%%% APPENDICES %%%%%%%%%%%%%%%%%%%%%
\appendix
\section{Validation on N-body simulations}
\label{sect:sims_buzzard}

\begin{figure*}
\includegraphics[width=1.\textwidth]{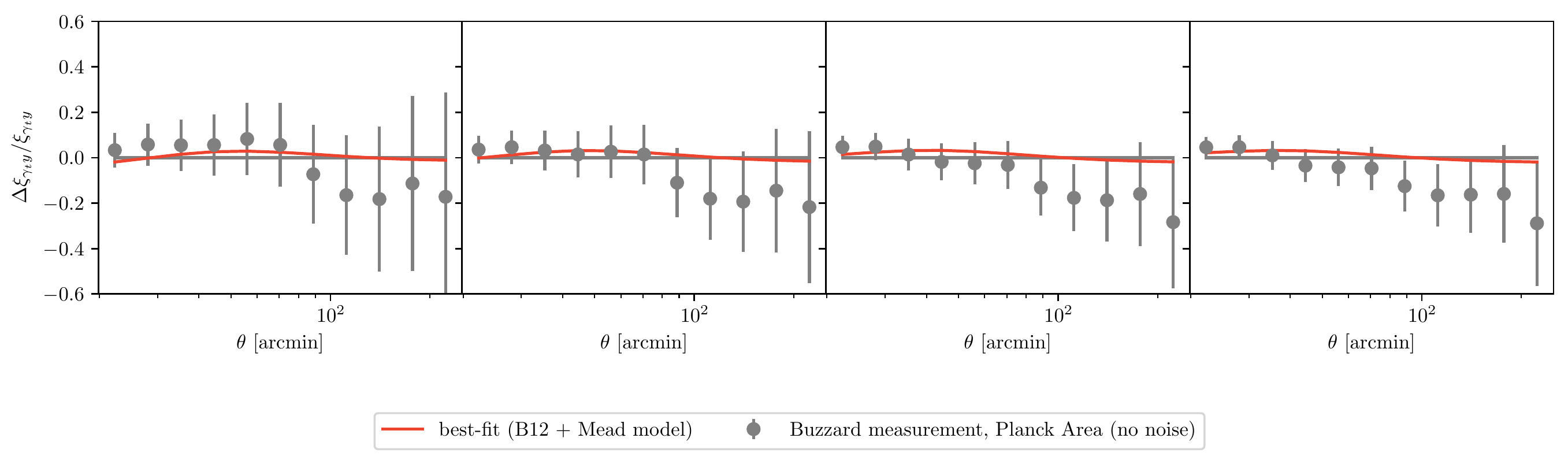}
\caption[]{Comparison between the measured  $\xi^{\gamma_t y}$ in Buzzard simulation and our theory predictions. In particular, we report the fractional difference with respect to the theory predictions. Theory predictions have been obtained assuming a B12 pressure profile and NFW profile for the DM profile.  Grey points represent the measurement on a unsmoothed, noise-free realisation of Buzzard. Uncertainties are estimated using a noise-free theoretical covariance. Measurement points are very correlated (both among different redshift bins and angular scales). The red line represents the best fit to the data performed using the Mead model instead of the NFW profile.}
\label{Buzzard_measurement}
\end{figure*}

We provide in this section further validation of our modelling by measuring the shear-Compton-$y$ map cross correlation on the fiducial DES Y3 N-body simulations. {Note that an independent modelling check on hydrodynamical simulations has been performed by \cite{Battaglia2015}, validating the use of the halo framework to model shear-Compton-$y$ cross-correlations.}

For this test, we use one realisation of the DES Y3 Buzzard catalogue v2.0 \citep{DeRose2018,y3-simvalidation}. Cosmological parameters of the simulation has been chosen to be  $\Omega_{\rm m} = 0.286$, $\sigma_8 = 0.82$, $\Omega_b = 0.047$, $n_s = 0.96 $, $h = 0.7$. The lightcone of the simulation is generated on the fly starting from three boxes with different resolutions and sizes (1050$^3$, 2600$^3$ and 4000$^3$ Mpc$^3 h^{-3}$ boxes and 1400$^3$, 2048$^3$ and 2048$^3$ particles); halos are identified using the public code ROCKSTAR \citep{Behroozi2013} and they are populated with galaxies using ADDGALS \citep{DeRose2018}. Lensing effects are calculated using the multiple plane ray-tracing algorithm CALCLENS \citep{Becker2013}. From the halo catalog, we construct a tSZ map by pasting a \cite{Battaglia_2012} profile on each halo. The map comes with in the \texttt{healpy} format with a resolution of \texttt{NSIDE} 4096. For this test, we do not smooth the map nor add instrumental noise. As for the simulated shape catalog, we use for this test a shape noise-free catalog, which faithfully reproduces DES Y3 area coverage and number density. Galaxies are further divided into four tomographic bins following the same methodology used on data.

We measure $\xi^{\gamma_t y}$ using the unsmoothed, noise-free simulated Compton-$y$ map over the \textit{Planck} footprint and the shape noise-free simulated shear catalog. As Buzzard is a DM-only N-body simulation, we model the signal using a NFW profile for the Fourier transform of the DM profile, rather than the Mead model. The comparison between the theory predictions and our measurements in simulations is shown in Fig.~\ref{Buzzard_measurement}. We only show the result of this comparison for scales larger than 20 arcminutes. Below such a scale, the simulation becomes unreliable, as the measurement points receive significant contributions from scales below the simulation resolution ($k = 3 h/$Mpc). In the scales where the comparison can be trusted, there is a very good match between the measurement and the theoretical predictions {(the slight offest at large scales for bin 3 and 4 is compatible with cosmic variance, which is captured by the error bars)}. In particular, for the four tomographic bins, we obtain a $\chi^2=3,4,4,5$ for 9 $d.o.f.$.

As a final test, we fitted our measurements using the Mead model, rather than the NFW profile, fixing all the parameters of our modelling except for the two Mead model parameters, which were sampled using broad flat priors. The Mead model encompasses the NFW profile as a subset of its parameter space, but it has additional flexibility. Indeed, the Mead model can provide a god fit for the Buzzard measurement at all scales.%, even if it is biased low at small scales due to resolution effect. This suggests that freeing the Mead model parameters is a pretty conservative choice, as they could absorb small-scales systematics or inaccuracies in our modelling. 

\section{Covariance Validation}\label{sec:covariance}

\begin{figure*}
\includegraphics[width=1.\textwidth]{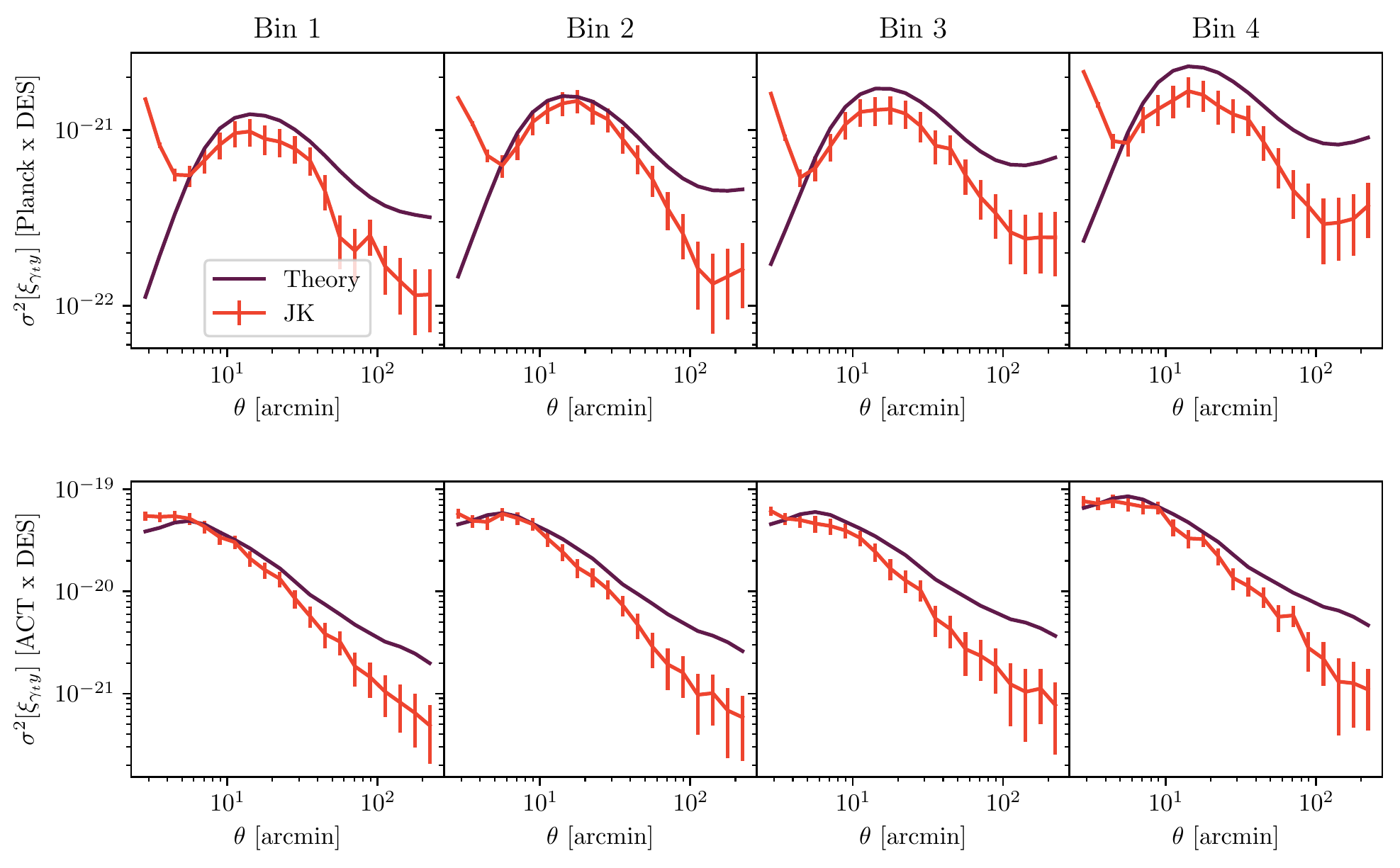}
\caption[]{Diagonal elements of the covariance matrix, for the \textit{Planck} and DES $\xi^{\gamma_t y}(\theta)$ (top panels) and ACT and DES $\xi^{\gamma_t y}(\theta)$ (lower panels), for the four different DES tomographic bins. In each panel, we compare theory predictions (purple) to jackknife estimates (JK, red).}
\label{cov_comparison1}
\end{figure*}

\begin{figure}

\includegraphics[width=0.4\textwidth]{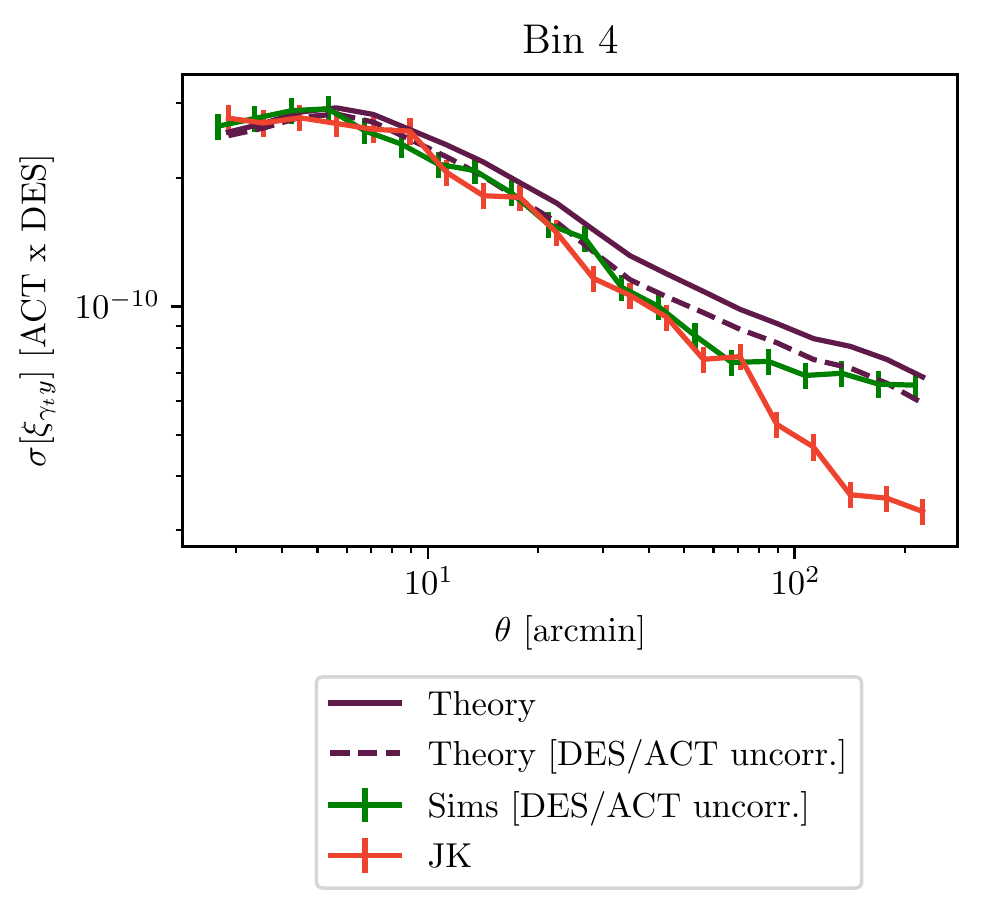}
\caption[]{Diagonal elements of the covariance matrix, for the fourth tomographic bin, for the ACTxDES $\xi^{\gamma_t y}(\theta)$. In addition to theory predictions (purple) and jackknife estimates (JK, red), we also show the estimates obtained by using ACT simulations and DES data (green), and the theory predictions that neglect correlations between ACT and DES (purple, dashed lines).}
\label{cov_comparison2}
\end{figure}

We model the covariance $\varmathbb{C}$ of the convergence and Compton-$y$ cross-spectra as a sum of Gaussian ($\varmathbb{C}^{\rm G}$) and non-Gaussian ($\varmathbb{C}^{\rm NG}$) terms as follows:
\begin{equation}
    \label{eq:cov_tot}
\varmathbb{C} (C^{\rm \kappa,y_i}_{\ell_1},C^{\rm \kappa,y_j}_{\ell_2}) = \varmathbb{C}^{\rm G} (C^{\rm \kappa,y_i}_{\ell_1},C^{\rm \kappa,y_j}_{\ell_2}) + \varmathbb{C}^{\rm NG} (C^{\rm  \kappa,y_i}_{\ell_1},C^{\rm  \kappa,y_j}_{\ell_2}),
\end{equation}
where $\kappa$ refers to the DES convergence field and $y_i$, $y_j$ represent either the \textit{Planck} or ACT Compton-$y$ field. The Gaussian term is given by \citep{Hu_2004}:
\begin{equation}
\label{eq:cov_g}
\varmathbb{C}^{\rm G} (C^{\rm \kappa,y_i}_{\ell_1},C^{\rm \kappa,y_i}_{\ell_2}) = \frac{\delta_{\ell_1 \ell_2}}{f^{\rm \kappa,y_i ; \kappa,y_j}_{\rm sky} (2 \ell_1 + 1)\Delta \ell_1} \bigg[ \hat{C}^{\rm \kappa,\kappa}_{\ell_1} \hat{C}^{\rm y_i,y_j}_{\ell_2} + \hat{C}^{\rm \kappa,y_j}_{\ell_1} \hat{C}^{\rm \kappa,y_i}_{\ell_2}  \bigg].
\end{equation}
Here, $\delta_{\ell_1 \ell_2}$ is the Kronecker delta, ${f^{\rm \kappa,y_i ; \kappa,y_j}_{\rm sky}}$ is the effective sky coverage fraction, $\Delta \ell_1$ is the size of the multipole bin, and $\hat{C}_{\ell}$ is the total cross-spectrum between any pair of fields including the noise contribution. The non-Gaussian part, can be written following \cite{Makiya2018}:
\begin{equation}
\varmathbb{C}^{\rm NG} (C^{\rm  \kappa,y_i}_{\ell_1},C^{\rm  \kappa,y_j}_{\ell_2}) = \frac{1}{4 \pi f^{\rm \kappa,y_i ; \kappa,y_j}_{\rm sky}} T^{\kappa,y_i\kappa,y_j}_{\ell_1\ell_2},
\end{equation}
with
\begin{equation}
T^{\kappa,y_i\kappa,y_j}_{\ell_1\ell_2} = \int dz \frac{dV}{dz d\Omega} \int dM \frac{dn}{dM} \bar{\kappa}_{\ell_1} \bar{y}_{i,\ell_1} \bar{\kappa}_{\ell_2} \bar{y}_{j,\ell_2}.
\end{equation}

The real-space covariance for the measurement is then obtained by:
\begin{multline}
\varmathbb{C}(\xi^{\gamma_t y_i}(\theta_1),\xi^{\gamma_t y_j}(\theta_2)) = \int \frac{d\ell_1 \ \ell_1}{2\pi} J_{2}(\ell_1 \theta_1) \times \\\int \frac{d\ell_2 \ \ell_2}{2\pi} J_{2}(\ell_2 \theta_2)  \varmathbb{C} (C^{\rm \kappa,y_i}_{\ell_1},C^{\rm \kappa,y_j}_{\ell_2}).
\end{multline}
In order to validate the covariance matrix, we follow \cite{Makiya2018} and perform a comparison with a covariance matrix estimated through jackknife resampling of the measurement on data. In particular, we use the following expression \citep{Norberg2009}:
\begin{equation}
\hat{\Sigma}(x_i,x_j)=\frac{(N_{\rm JK}-1)}{N_{\rm JK}} \sum_{k=1}^{N_{\rm JK}} (x_i^k-\bar{x_i})(x_j^k-\bar{x_j}),
\end{equation}
where the sample is divided into $N_{\rm JK}=200$ sub-regions of roughly equal area,  $x_i$ is a measure of the statistic of interest in the i-th bin of the k-th sample, and $\bar{x_i}$ is the mean of our resamplings. Note that the jackknife resampling only allows to efficiently estimate the covariance matrix on scales smaller than the size of the jackknife patches, so the jackknife covariance will be biased low at large scales.
The comparison between the theoretical and the jaccknife covariance for the $\xi^{\gamma_t y}$ measurements is shown in Fig.~\ref{cov_comparison1}, and shows good agreement at small-intermediate scales, where the jackknife covariance can be considered reliable. Note that the range of scales where this comparison holds is smaller for ACT, since average size of the JK patch is much smaller than in the case of \textit{Planck}. In the case of \textit{Planck}xDES covariance, the jackknife estimates have an upturn for scales smaller than 5 arcminutes which is not captured by our analytical covariance. Our guess is that this is related to mask effects; however, we did not investigate this further, as in the case of \textit{Planck}xDES measurement we exclude scales $<8$ arcminutes due to \textit{Planck} beam FWHM.

The $\xi^{\gamma_t y}$ measurement involving DES and ACT map is also validated cross-correlating 300 simulated ACT maps with the DES Y3 shape catalog. These measurements should capture the dominant part of the Gaussian part of the covariance ($\propto \hat{C}^{\rm \kappa,\kappa}_{\ell_1} \hat{C}^{\rm y_i,y_j}_{\ell_2}$), but they cannot capture the terms $\hat{C}^{\rm \kappa,y_j}_{\ell_1} \hat{C}^{\rm \kappa,y_i}_{\ell_2}$ and the non gaussian part of the covariance, since the large-scale fluctuations of the simulated Compton y-maps are not correlated with the data one from the shear catalog. This is still a relevant comparison as these two latter terms should not dominate in the case of the ACTxDES measurement. 

However, the non Gaussian part of the covariance is expected to be sub-dominant when cross correlating ACT with DES, so this should not strongly affect the comparison. Fig.~\ref{cov_comparison2} shows the diagonal elements of the covariance matrix estimated using ACT simulations, showing a better match with theory at large scales compared to jackknife estimates. For comparison purposes, we also show the theory covariance matrix computed dropping terms that are not captured by the measurement in simulations.

%In order to validate \textit{Planck} covariance we cannot use the same simulations used for ACT, as they only cover the ACT footprint. 

%- cross correlating simulated 400 act maps with des y3 data should capture the %gaussian part, modulo cl_ACT,DES that are negligible
%- 
%
%\begin{itemize}
%    \item non gaussian contribution is important only for planck
%    \item cross correlating simulated 400 act maps with des y3 data should %capture the gaussian part, modulo cl_ACT,DES that are negligible
%    \item for planck, we don't have analogous sims, we can only do JK, that are %valid at < intermediate scales. Shape noise dominates at small scales, so no %further test is possible
%\end{itemize}
\section{Tests on Websky mocks on CIB contamination}\label{sect:weird_sect}

\begin{figure}
\includegraphics[width=0.45\textwidth]{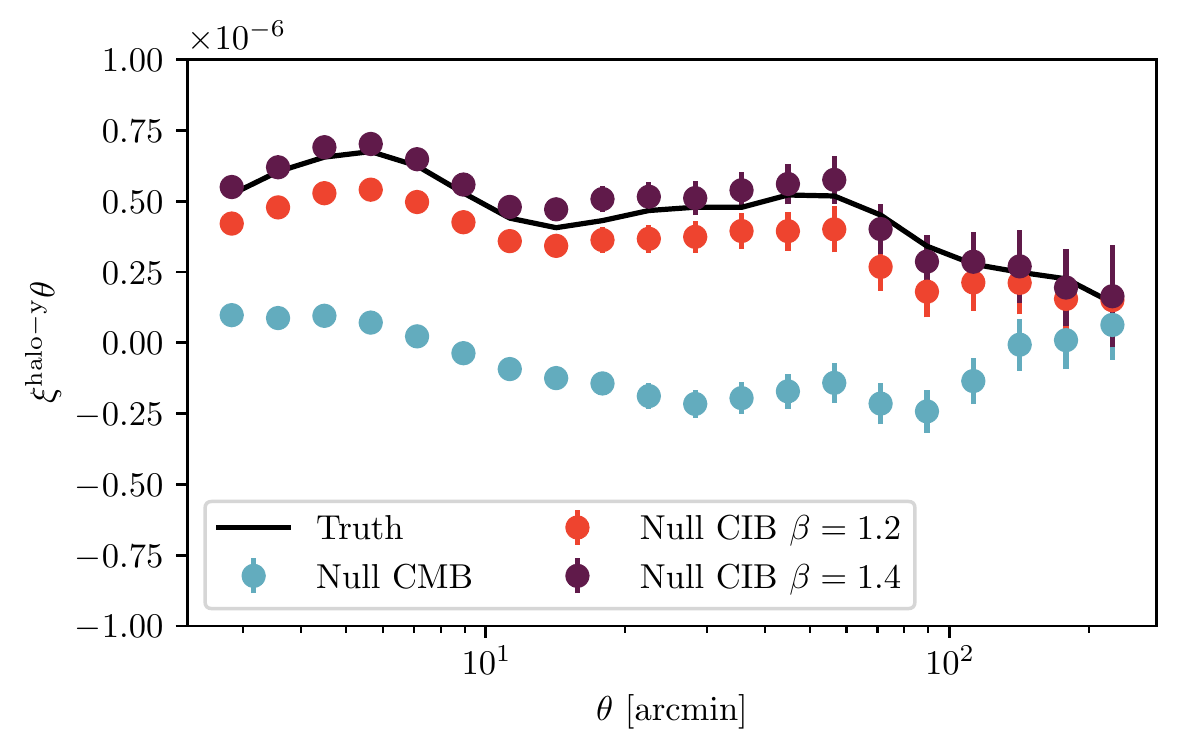}
\caption[]{Halo-Compton-$y$ correlation from Websky mocks. The three different measurements use three different versions of the Compton-$y$ maps. The black line refers to the case where the Compton-$y$ map is created starting from frequency channels without CIB contamination, whereas the two other measurements have been obtained using frequency channels contaminated by a fiducial CIB signal, with or without CIB de-projection at the map-making level. 
}
\label{weeird_plot}
\end{figure}

We further discuss in the Appendix the impact that the CIB can have on our measurements when the Compton-$y$ maps are not generated by explicitly de-projecting the CIB signal. {In our analysis, this has proven to be necessary for the \textit{Planck} x DES measurement, whereas we found no significant CIB contamination of the ACT x DES signal, owing to the lower signal-to-noise of the latter.} To this aim, we use Websky mocks \citep{Stein2019}, which are full-sky simulations of the extragalactic microwave sky generated using the mass-Peak Patch
approach. We use Compton-y, lensed CMB and CIB maps for frequencies 143, 217, 353, and 545 GHz, convolved with the nominal \textit{Planck} Gaussian beam and with \textit{Planck}-like white noise \citep{Planck_HIGH}. We created two sets of maps for each frequency channel: one with CIB contamination, and one without CIB contamination. Last, we created three Compton-$y$ maps using our NILC algorithm: a first map from the frequency channels without CIB contamination, a second map from the frequency channels with CIB contamination but without explicitly de-projecting the CIB signal, and a third one using the CIB contaminated frequency maps and de-projecting the CIB signal during the map making process. When de-projecting the CIB signal, we used $\beta=1.2$. The Websky mocks do not have shear maps available, but they provide a dark matter halo catalog. To qualitatively span the same redshift and halo mass range probed by our measurement, we selected halos so as to have a sample with average mean redshift $\avg{z} \sim 0.25$ and average halo mass $10^{14.3}\rm{M}_\odot$. This  corresponds to the typical redshift  and halo mass of our measurement involving the 4th tomographic bin (see \paperB). We then computed the halo Compton-$y$ correlation signal (obtained cross correlating the halos positions with the values of the Compton-$y$ maps). We show in Fig.~\ref{weeird_plot} the measurements with the three different Compton-$y$ maps. The angular scale sensitivity of this halo-Compton-$y$ correlation is expected to be different from the shear-Compton-$y$ correlation; moreover, when creating the simulated Compton-$y$ maps we did not use any frequency channels below 143 GHz, contrary to the maps on data. For these reasons, the effect of CIB on these measurements cannot be directly compared to the effect of CIB we see on data. Nonetheless, from Fig.~\ref{weeird_plot} is clear that if no CIB de-projection is implemented when making the Compton-$y$  map, the resulting measurement can be strongly biased.

\section{{Tests of feedback models} using DES prior}\label{sec:DES_prior}

\begin{figure}
\includegraphics[width=0.45\textwidth]{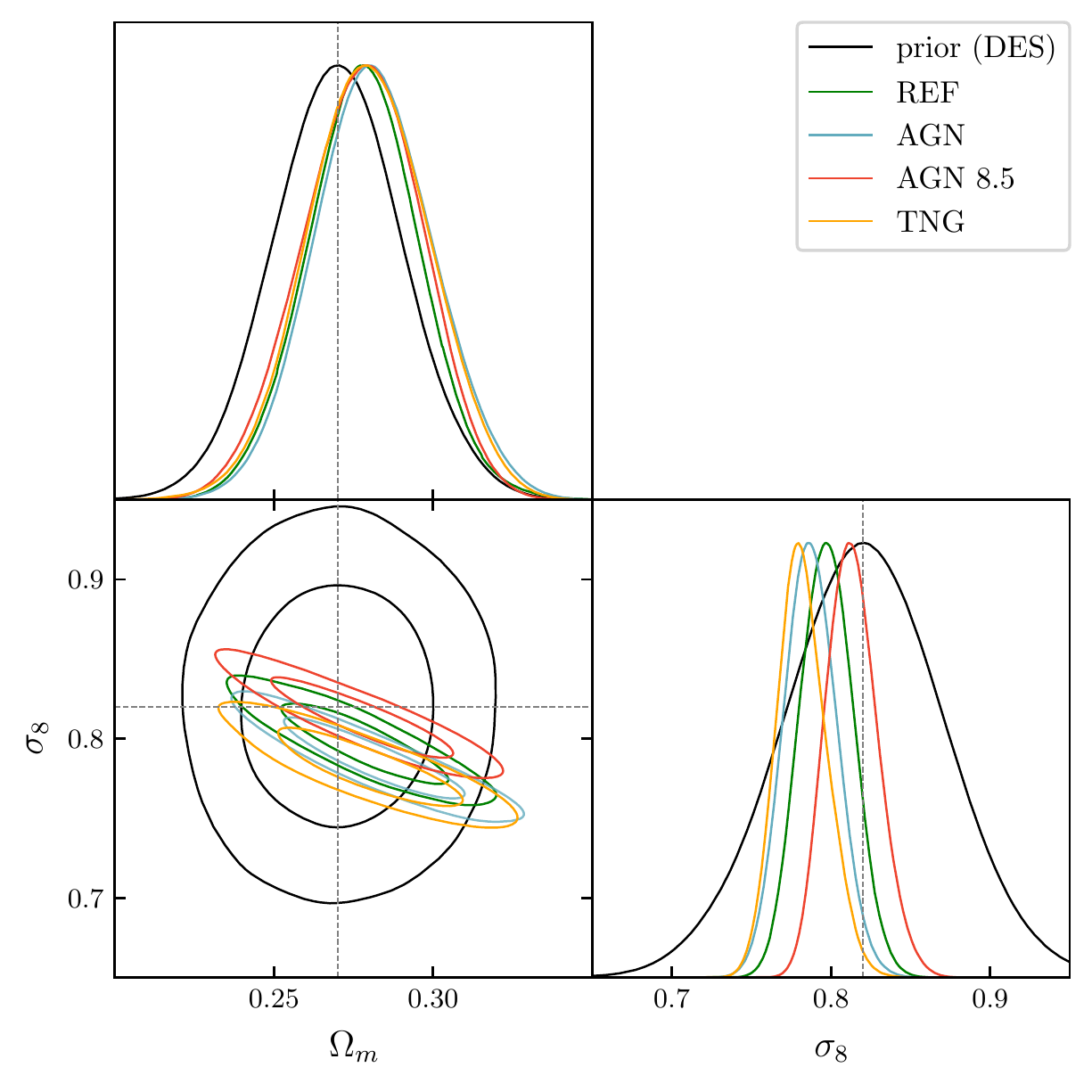}
\includegraphics[width=0.45\textwidth]{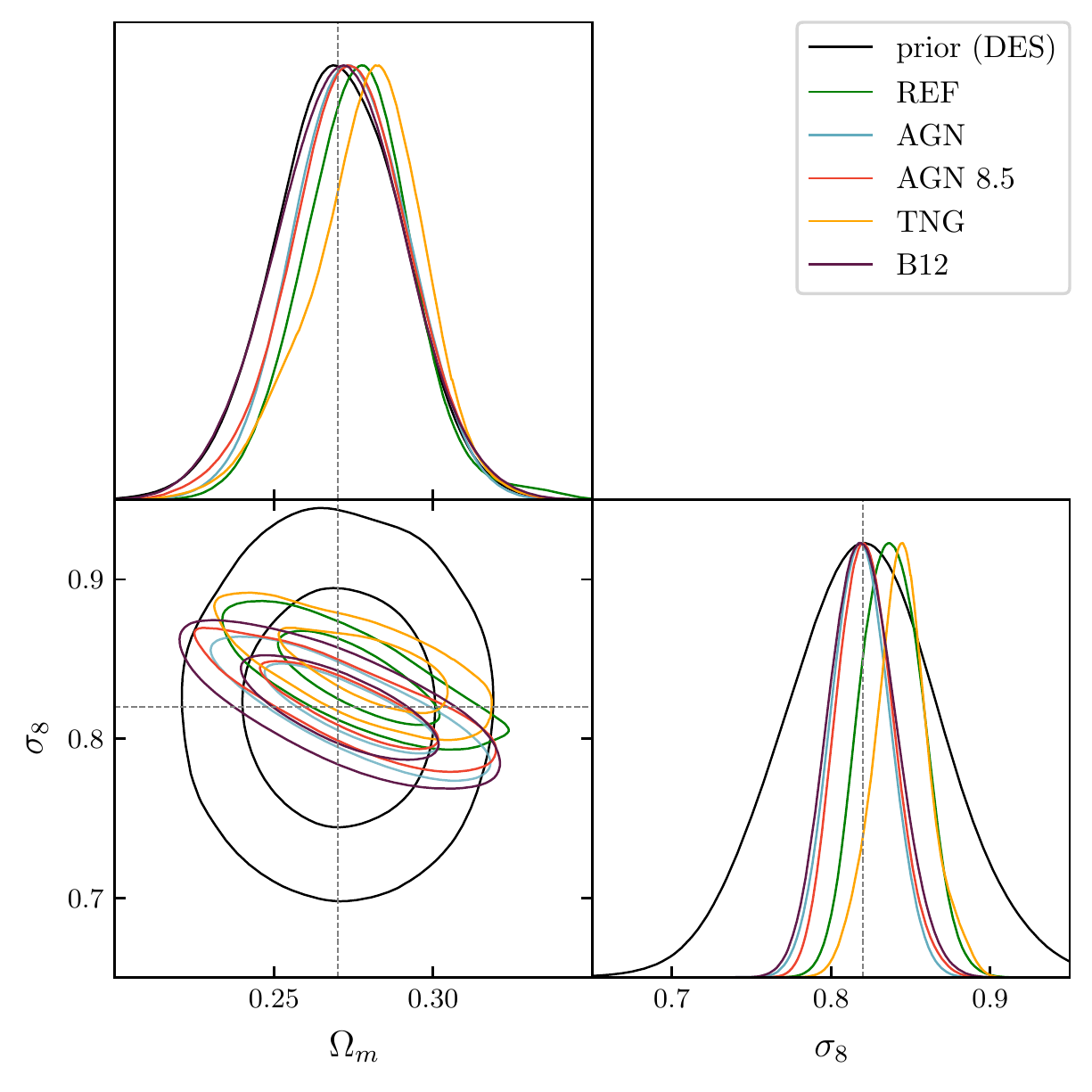}

\caption[]{Posterior for the parameters $\sigma_8$ and $\Omega_m$ for the four (five) feedback models: (B12) , REF, AGN, AGN 8.5 and TNG. We also show the \textit{DES} prior. Top refers to the case where we used the NFW re-scaling to model the lensing part of our signal; the bottom plot refers to the more conservative analysis where we used the Mead model.}
\label{contour_plots_DES}
\end{figure}

\begin{table}
%\tiny
\caption {Best fit $\chi^2$ for the four feedback models (B12, AGN, AGN8.5, REF), obtained assuming \textit{Planck} on $\sigma_8$ and $\Omega_{\rm m}$, and marginalising over nuisance parameters as explained in \S~\ref{sect:like}. The top table refers to the models obtained re-scaling the NFW profile for the lensing signal; on the other hand, the bottom table refers to the analysis where we model the lensing signal using the Mead model. We also report the update-difference-in-mean (UDM) tension for the best fit models with respect to their priors.}
\centering
%\begin{adjustbox}{width=0.7\textwidth}

\begin{tabular}{|c|c|c|c|c|c|}
 \hline
  \multicolumn{6}{|c|}{\,\, \, \,\,    \,\,       DES prior (NFW re-scaling)}    \\
 \hline
 & \textbf{B12} & \textbf{AGN} & \textbf{AGN 8.5} & \textbf{REF} &  \textbf{TNG} \\

% \hline
%bin 1 & & 49 & 49 & 48 & & 44 & 47 & 46 & &22 \\
%bin 2 & & 61 & 67 & 61 & & 53 & 59 & 57 & &29 \\
%bin 3 & & 79 & 101 & 84 & & 72 & 72 & 83 & &35 \\
%bin 4 & & 70 & 97 & 77 & & 61 & 62 & 73 & &35 \\

%(191/122)
%(7.4 $\sigma$)

%planck_B12_AGN 154
%planck_LB_REF 155
%planck_LB_AGN 155
%planck_LB_AGN85 154
%planck_TNG 154

 \hline

$\chi^2/d.o.f.$ &   \,\,\,\,\,\,\,\,-\,\,\,\,\,\,\,\,  & 170/119 & 158/119 & 187/119 & 194/119 \\
UDM tension &  \,\,\,\,\,\,\,\,-\,\,\,\,\,\,\,\,  & 0.5 $\sigma$ & 0.1 $\sigma$ & 0.3 $\sigma$ &0.6 $\sigma$ \\
\hline
\end{tabular}

\begin{tabular}{|c|c|c|c|c|c|c|}
 \hline
  \multicolumn{6}{|c|}{\,\, \, \,\,    \,\,       DES prior (free $A_{\rm Mead}$,$\eta_{\rm Mead}$) }    \\
 \hline
 & \textbf{B12} & \textbf{AGN} & \textbf{AGN 8.5} & \textbf{REF}& \textbf{TNG} \\

% \hline
%bin 1 & & 49 & 49 & 48 & & 44 & 47 & 46 & &22 \\
%bin 2 & & 61 & 67 & 61 & & 53 & 59 & 57 & &29 \\
%bin 3 & & 79 & 101 & 84 & & 72 & 72 & 83 & &35 \\
%bin 4 & & 70 & 97 & 77 & & 61 & 62 & 73 & &35 \\
 \hline

$\chi^2/d.o.f.$ &   154/118 & 154/118 & 154/118 & 156/118 &154/118  \\
UDM tension &  0.7 $\sigma$ & 0.2 $\sigma$ & 0.3 $\sigma$ & 0.2 $\sigma$  & 053 $\sigma$  \\
\hline
\end{tabular}

%\end{adjustbox}
\label{significance_DES}
\end{table}
We show in this Appendix the constraints of feedback models obtained using DES priors on $\sigma_8$ and $\Omega_m$. Table~\ref{significance_DES} shows the best-fit $\chi^2$ and the update-difference-in-mean (UDM) tension for the different feedback scenarios, for both cases where we use the NFW re-scaling to model the lensing part of our signal and where we use the Mead model instead. The posteriors of $\sigma_8$ and $\Omega_m$ are shown in Fig. \ref{contour_plots_DES}. The main difference with respect to our analysis using \textit{Planck} prior (\S~\ref{sect:feedback_models}) concerns the UDM tension metric, which does not show any sign of tension with the DES prior owing to the broader prior from the DES analysis compared to \textit{Planck}.  Besides this, similar to the \textit{Planck} prior case, we find that when implementing the NFW re-scaling, the data prefers the AGN 8.5 scenarios, whereas when implementing the Mead model, we are not able to discriminate among different feedback scenarios, owing to the less constraining nature of this modelling choice.

%%%%%%%%%%%%%%%%%%%%%%%%%%%%%%%%%%%%%%%%%%%%%%%%%%

% Don't change these lines
%\bsp	% typesetting comment
\label{lastpage}
\bibliography{bibliography}
\bibliographystyle{mn2e_2author_arxiv_amp.bst}

%\section*{Affiliations}
%\<input{aff.tex}

\end{document}